\newif\ifarxiv
\arxivtrue

\documentclass[lettersize,journal]{IEEEtran}

\usepackage{cite}
\usepackage{amsmath,amssymb,amsfonts}
\usepackage{graphicx}
\usepackage{textcomp}
\usepackage{xcolor}
\usepackage{caption}
\usepackage{algorithm}
\usepackage[noend]{algpseudocode}
\usepackage{scalerel}
\usepackage{subcaption}
\usepackage[nolist,nohyperlinks]{acronym}
\usepackage{tabularx}

\usepackage{tikz, circuitikz}
\usetikzlibrary{arrows,decorations.pathmorphing,backgrounds,positioning,fit,petri,quotes}
\usepackage{import}
\usetikzlibrary{svg.path}
\definecolor{orcidlogocol}{HTML}{A6CE39}
\tikzset{
  orcidlogo/.pic={
    \fill[orcidlogocol] svg{M256,128c0,70.7-57.3,128-128,128C57.3,256,0,198.7,0,128C0,57.3,57.3,0,128,0C198.7,0,256,57.3,256,128z};
    \fill[white] svg{M86.3,186.2H70.9V79.1h15.4v48.4V186.2z}
                 svg{M108.9,79.1h41.6c39.6,0,57,28.3,57,53.6c0,27.5-21.5,53.6-56.8,53.6h-41.8V79.1z M124.3,172.4h24.5c34.9,0,42.9-26.5,42.9-39.7c0-21.5-13.7-39.7-43.7-39.7h-23.7V172.4z}
                 svg{M88.7,56.8c0,5.5-4.5,10.1-10.1,10.1c-5.6,0-10.1-4.6-10.1-10.1c0-5.6,4.5-10.1,10.1-10.1C84.2,46.7,88.7,51.3,88.7,56.8z};
  }
}
\newcommand\orcidicon[1]{\href{https://orcid.org/#1}{\mbox{\scalerel*{
\begin{tikzpicture}[yscale=-1,transform shape]
\pic{orcidlogo};
\end{tikzpicture}
}{|}}}}
\usepackage{hyperref}
\hypersetup{
colorlinks=true, %
linktoc=all,     %
linkcolor=black,  %
}

\def\BibTeX{{\rm B\kern-.05em{\sc i\kern-.025em b}\kern-.08em
    T\kern-.1667em\lower.7ex\hbox{E}\kern-.125emX}}

\title{Event-Based Simulation of Stochastic Memristive Devices for Neuromorphic Computing}

\author{\IEEEauthorblockN{Waleed El-Geresy\IEEEauthorrefmark{1} \orcidicon{0000-0002-4016-6078},~\IEEEmembership{Graduate Student Member, IEEE},
 Christos Papavassiliou\IEEEauthorrefmark{2} \orcidicon{0000-0002-8003-2146},~\IEEEmembership{Senior Member, IEEE}, and Deniz Gündüz\IEEEauthorrefmark{3} \orcidicon{0000-0002-7725-395X},~\IEEEmembership{Fellow, IEEE}}
 \thanks{Authors are affiliated with the Department of Electrical and Electronic Engineering, Imperial College London. Emails: [waleed.el-geresy15\IEEEauthorrefmark{1}, c.papavas\IEEEauthorrefmark{2}, d.gunduz\IEEEauthorrefmark{3}] @imperial.ac.uk. This work was supported by the EPSRC under the DTP 2016-2017 (EP/N509486/1), DTP 2018-2019 (EP/R513052/1), SONATA (EP/W035960/1) and FORTE (EP/R024642/1) projects.
 \ifarxiv
 \else
 For the purpose of open access, the authors have applied a Creative Commons Attribution (CCBY) license to any Author Accepted Manuscript version arising from this submission.
 \fi
 }}

\begin{document}

\maketitle

\thispagestyle{plain}
\pagestyle{plain}

\begin{abstract}
    In this paper, we build a general modelling framework for memristors, suitable for the simulation of event-based systems such as hardware spiking neural networks, and more generally, neuromorphic computing systems composed of three independent components: i) an event-based modelling approach, extending and generalising an existing general model of memristors - the \ac{GMSM} \cite{molterGeneralizedMetastableSwitch2016} - eliminating errors associated with discrete time approximation, as well as offering potential improvements in terms of suitability for neuromorphic memristive system simulations; ii) a volatility state variable to allow for the unified understanding of disparate non-linear and volatile phenomena, including state relaxation, structural disruption, Joule heating, and non-linear drift in different memristive devices; and iii) a readout equation that separates the latent state variable evolution from explicit variables of interest such as an instantaneous resistance.
    We exhibit an illustrative implementation of this framework, fit to a resistive drift dataset for titanium dioxide memristors, based on a proposed linear conductance model for resistive drift in the devices.
    Finally, we highlight the application of the model to neuromorphic computing, through demonstrating the contribution of the volatility state variable to switching dynamics, resulting in frequency-dependent switching (for stable memristors acting as programmable synaptic weights) and the generation of action potentials (for unstable memristors, acting as spike-generators).
\end{abstract}


\begin{acronym}[] %
        \acro{AI}{Artificial Intelligence}

        \acro{ECM}{Electrochemical Metallization}
        
        \acro{GMSM}{Generalised Metastable Switch Model}
        \acro{GST}{Germanium Antimony Tellurium}

        \acro{LTP}{Long Term Potentiation}

        \acro{MIM}{Metal-Insulator-Metal}
        \acro{MSE}{Mean Squared Error}

        \acro{PCM}{Phase Change Memory}

        \acro{RRAM}{Resistive Random Access Memory}

        \acro{STDP}{Spike Timing Dependent Plasticity}
        \acro{STP}{Short Term Potentiation}
        \acro{STT}{Spin Transfer Torque}

        \acro{TCM}{Thermochemical Switching Mechanism}

        \acro{VCM}{Valence Change Mechanism}
\end{acronym}

\acresetall

\section{Introduction}
\begin{figure}
    \centering
    \includegraphics[width=\linewidth]{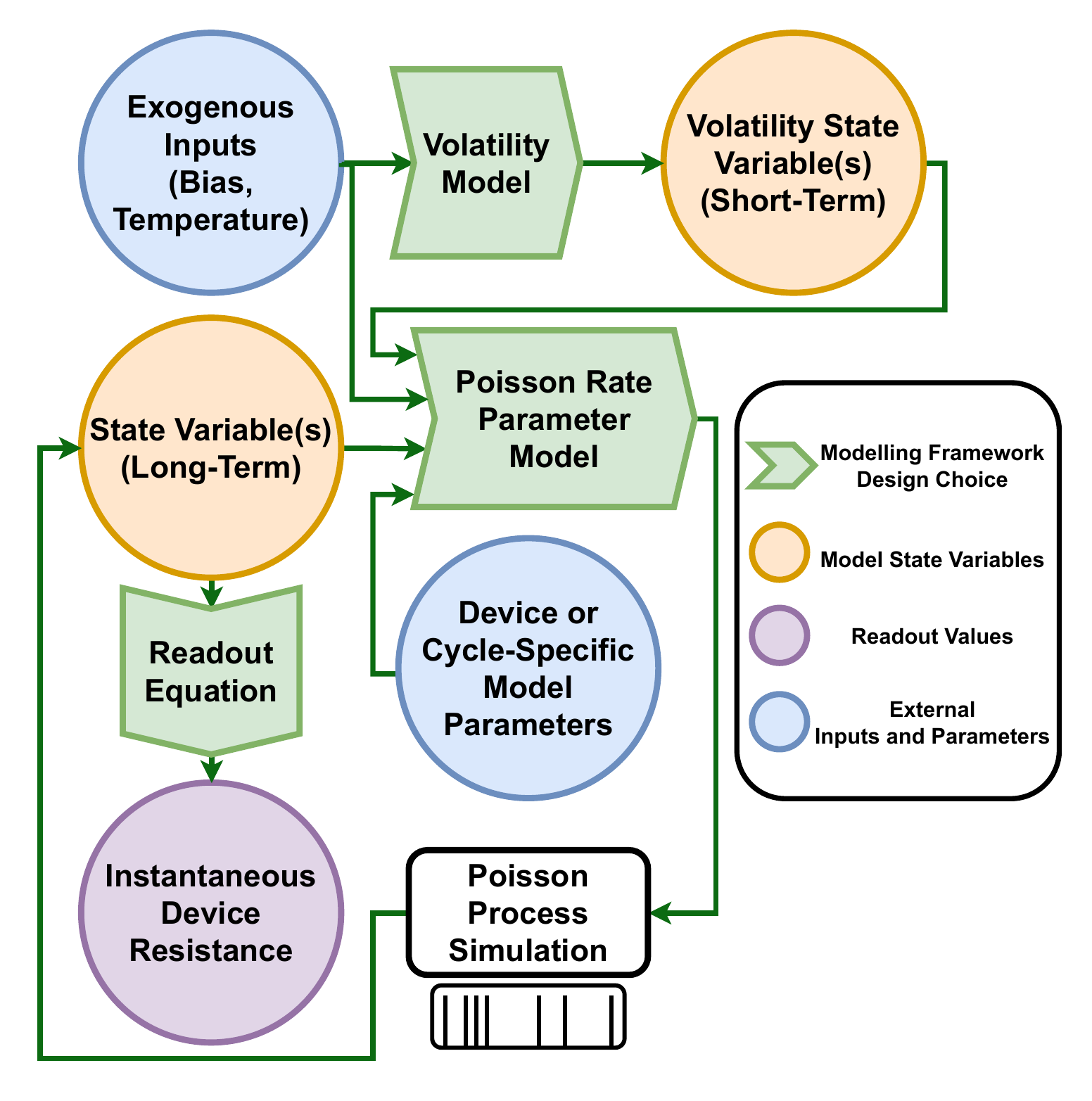}
    \caption{The event-based modelling framework. The framework consists of three components that are design choices: the Poisson rate equation (for state transitions), the volatility equation(s) (linking the exogenous inputs to the volatility state variable(s)), and the readout equation (that maps an internal state to a visible quantity such as the resistance).}
    \label{fig:neuromorphic_simulation}
\end{figure}

\IEEEPARstart{M}{emristors} are emerging passive electronic devices whose resistance can be both read and modified through the application of a voltage \cite{chuaMemristorMissingCircuit1971, chuaMemristiveDevicesSystems1976, chuaResistanceSwitchingMemories2011}. Growing research interest has been in part fuelled by their possible use as fundamental computational units in the field of neuromorphic computing - a computing paradigm that aims to take computational principles from neuroscience to improve the processing of information \cite{meadNeuromorphicElectronicSystems1990}.

The modelling of memristive devices is an important component of the development of novel architectures and systems based on these devices. Models can be classified along several dimensions: their generality, whether they are data-driven or physics based (or both), the inclusion of stochastic or non-ideal behaviour, and suitability for simulation in terms of their computational efficiency (compactness) and simplicity of description in the relevant simulation context. The last of these dimensions can be thought of as key - since the goal of a model is to enable the prediction of behaviour so as to allow for the design of strategies to utilise the behaviour of the devices to our advantage. Many deterministic memristor models have been proposed which focus on computational efficiency and compatibility with traditional electronic design software such as SPICE. However, modelling requirements for neuromorphic computing applications differ significantly. Notably, stochastic behaviour should be taken into account, modelling should be natively compatible with event-driven simulation, and the modelling approach should be naturally aligned with the design of neuromorphic systems.

In this paper, our goal is to develop a modelling framework for memristors that is suited for their use in neuromorphic computing applications. We firstly aim to extend existing modelling approaches for memristive devices to an event-based setting to allow the models to be used for event-driven simulations, such as for spiking neural networks. Secondly, we propose a simple and unified approach to modelling device volatility, and demonstrate how such an approach can provide a useful perspective for understanding existing and potential neuromorphic behaviours (such as action potential generation and potentiation) of memristive devices, through connecting the proposed \textit{volatility state variable} to the phenomena of short and long term potentiation.

The paper is structured as follows: in Section~\ref{sec:gmsm_background}, we begin by introducing an existing stochastic memristive model - the \ac{GMSM} \cite{molterGeneralizedMetastableSwitch2016, nugentAHaHComputingFromMetastable2014} and the state modelling approach. We introduce our modelling framework in Section~\ref{sec:event_based_metastable_switching_model}, composed of three main components: event-driven simulation, based on the evaluation of time-to-fire for multiple parallel Poisson point processes; volatility state variables as a way to generalise and unify the description of a variety of nonlinear, temporary and autonomous evolution phenomena observed in memristors; and the notion of the readout equation to allow for the separation of state variables from visible device readings such as resistance.
Following the introduction of the modelling framework, in Section~\ref{sec:metal_oxide}, 
we demonstrate an example implementation, applied to titanium dioxide memristors.
We present a drift dataset collected for the titanium oxide memristors, and demonstrate the fit of the model parameters to the observed data.
In Section~\ref{sec:accuracy_and_efficiency}, we offer a discussion of event-based component of the modelling approach, evaluating it qualitatively and comparing it to discrete timestep approximations.
Finally, in Section~\ref{sec:neuromorphic}, we provide demonstrations of ways in which the proposed modelling framework can be used in neuromorphic simulations. We introduce the notion of stable and unstable memristors, and give illustrative examples of ways in which neuromorphic behaviours - namely, frequency-dependent switching for stable memristors and the emulation of action potentials for unstable memristors - can be brought about through an understanding of the effects of volatility on switching speed and the equilibrium point, and the subsequent appropriate selection of modelling parameters. The source code for our model and simulations is made publicly available online \cite{el-geresySdpenguinMemristorneuromorphicmodel2025}.

\textbf{Notation} We use parameters \(E_x\) to denote the electronic energy barrier height, where \(V_x = E_x / q\) is the corresponding equivalent potential difference. We use the notation \(\mathcal{X}[i]\) to denote the \(i\)'th element of an ordered set \(\mathcal{X}\).

\section{Background}
\label{sec:gmsm_background}
\begin{figure}
    \centering
    \includegraphics[width=\linewidth]{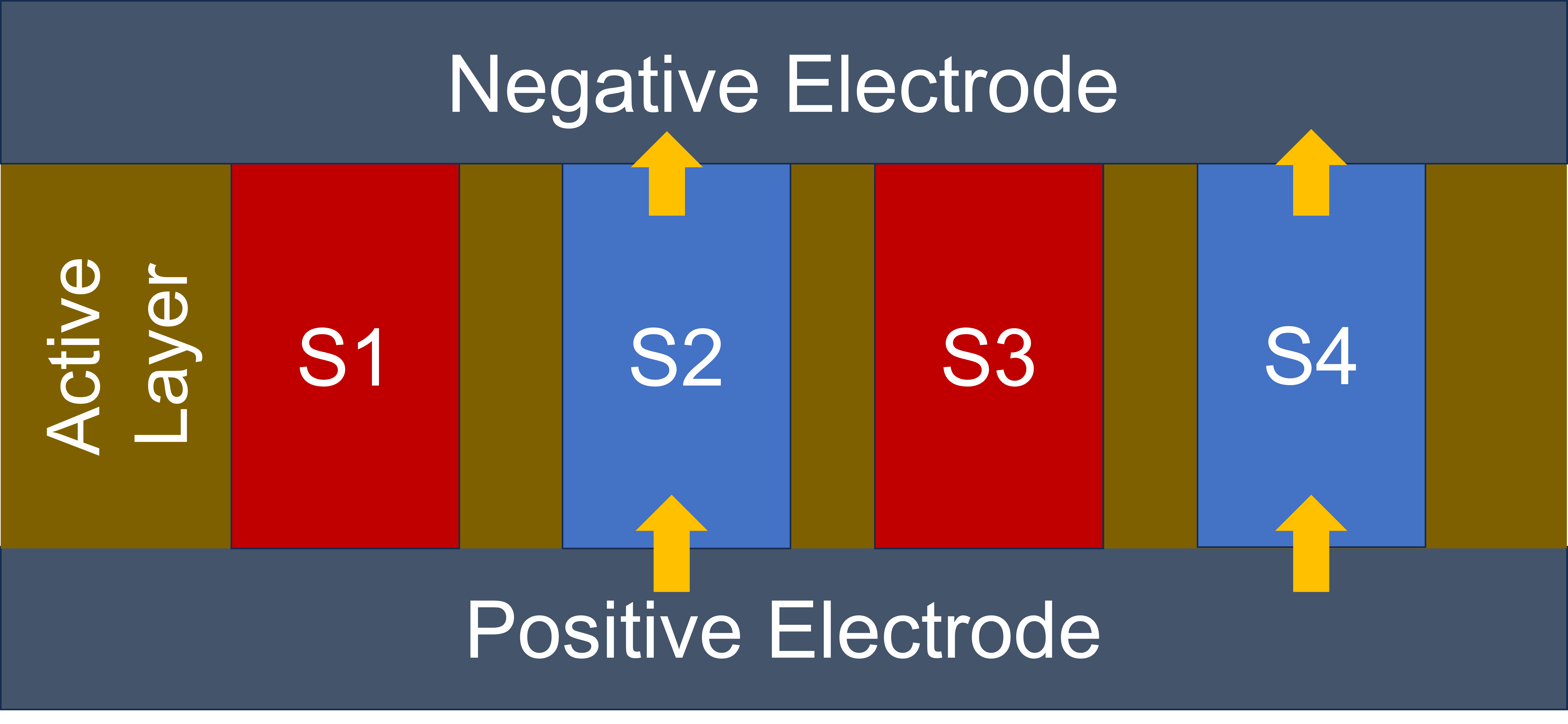}
    \caption{The metastable switch model. The conductance is a linear function of the number of switches (parallel conducting channels) that are in the ``on'' state (blue - S2 and S4 above) at any one time, as opposed to those in the ``off'' state (red - S1 and S3 above).}
    \label{fig:metastable_switches}
\end{figure}

Many models for memristive devices have been proposed; however, only a handful of these take into account the stochasticity inherent in the devices \cite{malikStochasticCompactModel2021, naousStochasticityModelingMemristors2016}, with the majority making deterministic assumptions.
One general memristor model that includes stochasticity as a fundamental feature of the model, is Molter \& Nugent's \ac{GMSM} \cite{molterGeneralizedMetastableSwitch2016}. It models the memristor as a number of parallel metastable switches, each allowed to take one of two different states - the ``high resistance'' or ``low resistance'' state (see Figure~\ref{fig:metastable_switches}), with stochastic transitions occurring every unit time with probabilities that are dependent on the temperature and bias.

The \ac{GMSM} bases its switching mechanism upon a Fermi-Dirac-like activation function (a logistic function), based on the temperature and the bias, which dictates the rate at which switching events occur. Two switching processes are modelled: low resistance to high resistance, and high resistance to low resistance. The probability of a single metastable switch transitioning from the low resistance state to the high resistance state in a given timestep \(h\) is denoted by \(P_A\), while the probability of a transition from the high resistance state to a low resistance state is denoted by \(P_B\) . These two probabilities are given by

\begin{align}
    P_A &= \alpha \frac{1}{1+e^{\beta(V-V_A)}},
    \label{eq:molter_model} \\
    P_B &= \alpha \frac{1}{1+e^{\beta(-V+V_B)}},
    \label{eq:molter_model2}
\end{align}
where \(V\) is the applied bias, \(V_A\) and \(V_B\) can be thought of as energy barrier heights that modulate the switching probabilities, \(\beta \triangleq \frac{q}{k_BT}\) is a physical parameter that introduces temperature-dependence and is the inverse of the thermal voltage (\(k_B\) is Boltzmann's constant and \(q\) is the electronic charge), and \(\alpha=\frac{h}{t_c}\) is a ratio of the timestep to the characteristic timescale \(t_c\) - a parameter of the device.

The \ac{GMSM} models the device's state evolution over time by considering the number of switches in either the high or low state and using the respective transition probabilities to define two binomial distributions that can be used to simulate the number of respective transitions of switches from low resistance to high, or vice versa for the second distribution, over a time-step. The normal approximation of binomial random variables can be used, as \(N\) becomes large, to form a continuous stochastic model.

In keeping with the terminology of the \ac{GMSM}, we use the terms ``switches'', or ``metastable switches'', to refer to the state variables of the model (the volatility is separately dealt with). For the purposes of this work, we assume that both the number of state variables (\(N\)) and the multiplicity of the individual state spaces are finite. Specifically, we consider the case where \(k=2\) - i.e. binary switches. Thus, the multiplicity of the state space (number of possible memristive states) is \(2^N\).
We use \(n\) to denote the number of switches in the state that is conducive to conduction, which we call the ``up state'', the ``high conductance state'', or the ``low resistance state''. Then, \(N - n\) denotes the number of switches in the state that is not conducive to conduction, which we call the ``down state'', the ``low conductance state'', or the ``high resistance state''. Switching from a lower to a higher resistance state is termed ``up'' switching, while switching in the opposite direction is termed ``down'' switching, reflecting the respective increase or decrease in the resistance.
It has been shown that a single state variable is sufficient to reasonably explain the behaviour of various memristive devices \cite{strukovMissingMemristorFound2008, pickettSwitchingDynamicsTitanium2009}.

\section{Modelling Framework}
\label{sec:event_based_metastable_switching_model}
In this section, we introduce the proposed modelling framework. Our framework consists of three modular components: a rate equation, volatility equation(s), and a readout equation, which can each be different depending on the memristive device being modelled. In Section~\ref{sec:switching_mechanism}, we start by introducing our adaption to the event-based setting, basing the simulation on continuous time probabilistic point processes, rather than discrete time Bernoulli random variables. Next, in Section~\ref{sec:volatility_state_variable}, we introduce the ``volatility state variable'' to allow for unified description of a range of physical phenomena including autonomous state evolution, Joule heating, and nonlinear drift, which in turn will allow for the understanding of certain short term memory behaviours in memristive devices (see Section~\ref{sec:neuromorphic}).  Section~\ref{sec:resistance_readout} introduces the readout equation to separate state variables from summary parameters of interest such as the instantaneous device resistance. In Section~\ref{sec:practical_simulations}, we specify the event-based simulation algorithm and a procedure for incorporating non-event-based inputs.

\subsection{Event-Based Switching and the Rate Parameter Model}
\label{sec:switching_mechanism}

As detailed, the \ac{GMSM} models the metastable switching dynamics through a set of \(N\) parallel Bernoulli random variables (one for each switch) for each timestep. The \(N\) independent Bernoulli random variables are then combined into two binomial random variables - or, in the continuous approximation, a normal distribution - that measure the state change for all switches in a single timestep \(\Delta T\). Motivated by the use of memristors in event-based neuromorphic computing systems, we adapt the modelling of the state transitions from a timestep-based transition probability to a (Poisson) point process, whose rate parameter dictates the frequency of switching events. This approach changes from a timestep-based evaluation to an event-based evaluation, where we model the changes to the device's state as and when they occur. This removes disctete-time simulation errors. It also offers potential benefits from a computational efficiency perspective, since the frequency of model updates are tied to frequency of change of the memristive state.

We replace the Bernoulli probabilities with Poisson rates for independent Poisson point processes. We use the convention that the subscript \textit{\(\text{up}\)} refers to parameters describing the up transition (and the same, respectively, for subscript \(\text{down}\) and down transitions).

The rate parameter for the Poisson process of a single metastable switch is referred to as the base rate and denoted by either \(\lambda'_{\text{up}}\) or \(\lambda'_{\text{down}}\).
To obtain the overall rate (the cumulative rate of all up or down switching events), we simply scale the base rate by the number of switches in the origin state - \(m\) for \(\lambda'_{\text{up}}\), or \(n\) for \(\lambda'_{\text{down}}\), giving us the cumulative up and down switching parameters, \(\lambda_{\text{up}}\) and \(\lambda_{\text{down}}\). We use the convention that a positive voltage makes it more probable for the device to switch to a higher resistive state, while a negative voltage makes it more probable for the device to switch to a lower resistive state.

The calculation of the Poisson rate parameters for the state transitions of the metastable switches in their given state(s) will be device and material-specific. The role of the Poisson rate parameter model is shown in Figure~\ref{fig:neuromorphic_simulation}. Device-to-device and cycle-to-cycle variability can be easily incorporated into the modelling by varying the parameters of the rate parameter model by device instance or programming cycle, respectively.

\subsection{Volatility State Variable}
\label{sec:volatility_state_variable}

Different types of memristive device differ significantly in their structure, material composition, and switching/conduction mechanics, but can nevertheless display similar characteristics. We hypothesise that a single state variable can be used to describe the effect of these different phenomena for neuromorphic computing applications.
Such characteristics include memory-dependence in switching and conduction behaviour; self-excitatory or self-inhibitory behaviour during switching and conduction leading to non-linear switching and/or conduction; and autonomous state evolution as a result of temporary changes within the devices, such as increased temperature or disorder in the material. Examples of phenomena that exemplify these characteristics include:

\begin{enumerate}
    \item Joule heating in metal oxide and \ac{PCM} devices.
    \item Structural relaxation in \ac{PCM} devices \cite{galloOverviewPhasechangeMemory2020, mantegazzaIncompleteFilamentCrystallization2010}.
    \item Interfacial energy minimisation in diffusive memristive devices \cite{wangMemristorsDiffusiveDynamics2017}.
\end{enumerate}

These phenomena are unified by their modulation of the switching characteristics under particular bias and temperature conditions. All of these phenomena can cause a temporary increase in the rate of change of the memristor state variable for a given voltage bias. This can manifest in the form of several switching phenomena:

\begin{enumerate}
    \item Metastability - the tendency for a device to revert to a more stable neighbouring state after switching.
    \item Resistive drift (autonomous state evolution) - a phenomenon involving a change of state under zero bias.
    \item Non-Linear switching e.g. exponential ionic drift \cite{strukovExponentialIonicDrift2009}.
    \item Accelerated switching behaviour, where the frequency or duration (spacing in time) of applied voltage pulses impacts the rate of switching (see Section~\ref{sec:frequency_dependent_potentiation}).
\end{enumerate}

As such, we propose a unified modelling approach in the form of a ``volatility state variable''. The term ``volatility'' is sometimes used, in the sense of instability, as a synonym for the phenomenon of resistive drift - the tendency for the resistance of a device to deviate from its initial value over time. We here use the term to instead refer to any kind of transient physical phenomenon that temporarily affects the rate of switching or conduction of a memristive device. This general description is of value from the perspective of modelling for neuromorphic computing applications, where we would like to understand the aforementioned behaviours such that we can control them to perform computation.

We define a volatility state variable, \(\rho(t)\), as follows: one or more state variables which directly modify the \textbf{global} rate of switching (both \(\lambda_{\text{up}}\) and \(\lambda_{\text{down}}\)), which has a characteristic time constant of decay, \(\tau_x\) and will tend to decay to an equilibrium value \(\rho_{\text{eq}}\), which may be modulated by external (exogenous) inputs of voltage or temperature:

\begin{align}
    \frac{d\rho(t)}{dt} = \left(\rho_{\text{eq}}(v(t), T(t)) - \rho(t)\right) / \tau_{\text{volatile}}
\end{align}

The exact nature of the volatility model will be dependent on the type of device. Its role is also visualised in Figure~\ref{fig:neuromorphic_simulation}.

\subsection{Readout Equation}
\label{sec:resistance_readout}

The voltage-current relationship for a memristive device need not always be instantaneously linear. Indeed, an exponential or \(\sinh\) relationship is possible depending on the nature of the device being considered \cite{molterGeneralizedMetastableSwitch2016}. Different behaviour can be engineered by relating the memristive state variable, \(n\), to a separate quantity such as the parameter for a Shockley diode equation, or the scaling factor for a \(\sinh\) voltage-current relation, depending on the device.

Rather than requiring that the resistance be linearly linked to the state variable, we can instead use a readout equation to provide flexibility in simulation. The readout equation takes as input the memristive state variable \(n\), and outputs a resistance. This allows for a consistent description of the state, and allows for a modular approach that separates state evolution dynamics from the resistance, which can be considered a summary statistic of a more complex state.
In Section~\ref{sec:metal_oxide}, we see an example of a thresholding, non-linear readout equation, that links the state variable to the voltage-current characteristics in a non-linear way. The role of the readout equation in the framework is visualised in Figure~\ref{fig:neuromorphic_simulation}.

\subsection{Practical Simulations}
\label{sec:practical_simulations}

Let us consider the switching of a single metastable switch to be modelled by a memoryless exponential distribution with the rate parameter \(\lambda\) proportional to the probability of switching. Note that this model is memoryless - the distribution for the evolution of the resistive state can be completely characterised by the current state, with no reference to the previous states of the device.

\subsubsection{Point Process Sampling}

Considering the cumulative distribution function for this exponential distribution, we can use inverse transform sampling (the Smirnov transformation) to generate samples for the point process:

\begin{align*}
p(T \leq t) &= 1 - e^{-\lambda t} \\
y &= 1 - e^{-\lambda t} \\
t &= -\frac{\ln{(1-y)}}{\lambda}
\end{align*}

We can perform the simulation of parallel point process sampling in two ways. We note
that we can sample the time from a point process with rate equal to the sum of all rates of the independent point processes, and subsequently label the switching events post-fact. Alternatively, we can sample times for a cumulative ``up``-switching process and a cumulative ``down''-switching process and take the smallest inter-arrival time as the next event, and repeatedly resample.

\subsubsection{Changes to the Distribution at Events}

When an event occurs (either a switching up or switching down event), this impacts the distributions and thus we need to reevaluate a sample for the switching time (for both the upwards and downwards processes).

We use the memoryless property of the exponential distribution, which states that if an event has not occurred by time \(\tau\) then the distribution for an event to occur at time \(t+\tau\) is still exponentially distributed. The only distinction is that we now modify the rate parameter for each distribution, which can be calculated as a simple linear scaling of the rate parameter, since the sum of N independent and identically distributed exponential distributions with rate parameter \(\lambda\) is an exponential distribution with rate parameter \(N\lambda\), as described in Section~\ref{sec:switching_mechanism}.

\subsubsection{Simulation Algorithm}
The procedure for running the model simulation is given in Algorithm~\ref{alg:switching_events_with_non_switching}. \(R\) corresponds to the array of generated resistances (with \(r_{\text{init}}\) being the initial resistance, determined by the initial state parameter \(n\)) and \(T\) corresponds to the array of times at which the resistance changes. \(\sim\) denotes sampling from a given probability distribution. \(f_{\text{ro}}\) is the readout equation that gives us a resistance value for a given state, as detailed in Equation~\ref{eq:resistance_readout}.

Note that in the case the value of \(n\) falls to zero, or reaches \(N\), we simply wait for an appropriate event to occur, to increase or decrease the value of \(n\), respectively. Algorithm~\ref{alg:switching_events_with_non_switching} shows the switching process for a total simulation time \(T_{\text{max}}\).

\paragraph{Non-Event-Based Inputs}
\label{sec:non_event_based_inputs}

In order to incorporate non-event based inputs, including variables that may be continuous functions of time such as temperature or applied bias, or piecewise constant functions, we can choose to apply one of two approaches.
If the values are unknown beforehand, we can include them as additional point processes with deterministic interarrival times equal to a fixed sampling timestep, \(t_{\text{sample}}\), which lower bounds the error in their values to that of a discrete time simulation, introducing a point process, \(\mathcal{T}_{\text{sample}}\), for updates, where e.g. a Euler integration update step is performed:

\begin{align}
    \mathcal{T}_{\text{sample}} = \{t_{\text{sample}}, 2t_{\text{sample}},3t_{\text{sample}} \ldots,\}
    \nonumber
\end{align}

However, if signal values are known beforehand then we can also sample using threshold-based, or event-driven, sampling.
Given an independent input signal \(m(t)\), we can define the associated marked point process, with the set of times \(\mathcal{T}\), at which the events occur, to be:

\begin{align*}
    \mathcal{T} &= \{t: m(t) \in \mathcal{M}\} \\
\end{align*}
where, for continuous functions the set \(\mathcal{M}\) contains the function values, changing in steps of \(\Delta m\):

\begin{align*}
    \mathcal{M}_i &= \{m(0), m(0) + \Delta m, \cdots , m(0) + l\Delta m\} \\ &\cup \{m(0), m(0) - \Delta m, \cdots , m(0) - l\Delta m\}
\end{align*}
i.e. a set of signal values at specified boundaries and their associated event times.
For piecewise constant (discontinuous) functions, \(\mathcal{M}\) will simply be the set of amplitudes at points when the function changes, \(\mathcal{T}\).
Due to the memoryless nature of the Poisson process, incorporating such sampling into the event-based modelling is straightforward - whenever \(t \in \mathcal{T}\), for a particular associated set of events, we run the event (update all continuous time variables and rate equations) and resample times to fire for the Poisson switching processes.

Although the method of conversion for continuous functions for inclusion in the event-based modelling framework works to minimise the error associated with changes to parameters affecting the rate equations not matching those used during the simulation - which are dependent purely on external stimuli - additional challenges are presented in the case of variables that are instead, or also, functions of the device state themselves. For example, volatility functions that update volatility state variables, such as Joule heating, will likely be dependent on the current through or voltage across the memristor. In this case, the variable must be re-evaluated whenever there is a change in the resistance, or in the voltage. Indeed, for rate equations that are functions of both external stimuli and the internal state variable, these should be re-evaluated every time any event happens, so as to avoid errors associated with the drift of the independent variables of the function away from their original values.
Let \(\{j_0 \cdots j_l\}\) be the update functions for the variables \(\{J_0 \cdots J_l\}\), each with initial values \(J_l^0\), which depend on at least one other variable for their updates and as such cannot be marked processes since they cannot be precomputed. A special marked process \(\mathcal{T}_{\text{sample}}\) is associated with this. These are updated whenever any other variable \(n\) is modified (i.e. an event occurs) by e.g. Euler integration.

\begin{algorithm}
	\caption{Event-based modelling procedure.}
	\label{alg:switching_events_with_non_switching}
	\begin{algorithmic}[1]
        \State $R \gets R_{\text{init}}$,  $n \gets n(R)$, $J_l \gets J_l^0$, $M_i \gets \mathcal{M}_i[0]$
        \State $\lambda'_{\text{up}} \gets f_{\text{up}}(\{M_i\})$, $\lambda'_{\text{down}} \gets f_{\text{down}}(\{M_i\})$
        \State $c_i \gets 0$, $t \gets 0$
        \State $F \gets 0$ \Comment{Denotes whether a firing event has occurred.}
        \While{$t \leq T_{\text{max}}$}
        \State $t_{\text{down}} \sim \text{Exp}((N-n) \cdot \lambda'_{\text{down}})$
        \State $t_{\text{up}} \sim \text{Exp}(n \cdot \lambda'_{\text{up}})$
        \ForAll{$i$} \Comment{State-independent firing.}
            \If{$F = 0 \; \textbf{and} \; c_i \leq |\mathcal{T}_{i}| \; \textbf{and} \; \mathcal{T}_{i}[c_i] - t\leq \{ t_{\text{sample}}, t_{\text{up}}, t_{\text{down}},  \mathcal{T}_{k}[c_k] - t \; \forall k \neq i \}$}
                \State $t_{\text{diff}} \gets \mathcal{T}_{i}[c_i] - t$
                \ForAll{$j$}
                
                    \State $J_l \gets j_l(t_{\text{diff}}, n, M_0, \cdots, M_i, J_0, \cdots, J_l)$
                \EndFor
                \State $M_i \gets \mathcal{M}_i[c_i]$
                \State $\lambda'_{\text{up}} \gets f_{\text{up}}(n, M_0, \cdots, M_i, J_0, \cdots, J_l)$
                \State $\lambda'_{\text{down}} \gets f_{\text{down}}(n, M_0, \cdots, M_i, J_0, \cdots, J_l)$
                \State $t \gets \mathcal{T}_{i}[c_i]$
                \State $c_i \gets c_i + 1$
                \State $F \gets 1$
            \EndIf
        \EndFor
        \If{$F = 0 \; \textbf{and} \; (t_{\text{down}} < t_{\text{sample}}) \; \textbf{and} \; ( t_{\text{down}} \leq t_{\text{up}} \; \textbf{or} \; n = N ) \; \textbf{and} \; n \neq 0 $}
            \ForAll{$j$} %
                \State $J_l \gets j_l(t_{\text{down}}, n, M_0, \cdots, M_i, J_0, \cdots, J_l)$
            \EndFor
            \State $n \gets n - 1$
            \State $\lambda'_{\text{up}} \gets f_{\text{up}}(n, M_0, \cdots, M_i, J_0, \cdots, J_l)$
            \State $\lambda'_{\text{down}} \gets f_{\text{down}}(n, M_0, \cdots, M_i, J_0, \cdots, J_l)$
            \State $t \gets t + t_{\text{down}}$
        \ElsIf{$F = 0 \; \textbf{and} \; (t_{\text{up}} < t_{\text{sample}}) \; \textbf{and} \; (t_{\text{up}} < t_{\text{down}} \; \textbf{or} \; n = 0)$}
            \ForAll{$j$}
                \State $J_l \gets j_l(t_{\text{up}}, n, M_0, \cdots, M_i, J_0, \cdots, J_l)$
            \EndFor
            \State $n \gets n + 1$
            \State $\lambda'_{\text{up}} \gets f_{\text{up}}(n, M_0, \cdots, M_i, J_0, \cdots, J_l)$
            \State $\lambda'_{\text{down}} \gets f_{\text{down}}(n, M_0, \cdots, M_i, J_0, \cdots, J_l)$
            \State $t \gets t + t_{\text{up}} $
        \ElsIf{$F=0$}
            \ForAll{$j$}
                \State $J_l \gets j_l(t_{\text{sample}}, n, M_0, \cdots, M_i, J_0, \cdots, J_l)$
            \EndFor
            \State $\lambda'_{\text{up}} \gets f_{\text{up}}(n, M_0, \cdots, M_i, J_0, \cdots, J_l)$
            \State $\lambda'_{\text{down}} \gets f_{\text{down}}(n, M_0, \cdots, M_i, J_0, \cdots, J_l)$
            \State $t \gets t + t_{\text{sample}} $
        \EndIf
        \State $R \gets f_{ro}(n) $
        \State $F \gets 0$
        \EndWhile
    \end{algorithmic}
\end{algorithm}

\paragraph{Conversion to Discrete Time}

If necessary, we can convert non-uniform series from the event-based model to uniform data in time through sampling, albeit introducing errors associated with the discretisation in time, as shown in Algorithm~\ref{alg:sampling}. \(\mathcal{T}_{switch}\) denotes the ordered set of switching times, and \(\mathcal{R}\) denotes the associated ordered set of values of a state variable at the times \(\mathcal{T}_{switch}\). \(T_{\text{sample}}\) denotes the sampling period and \(T_{\text{tot}}\) denotes the maximum desired time to sample.

\begin{algorithm}[t]
	\caption{Procedure for conversion of the non-uniform time series to uniformly sampled time series.}
	\label{alg:sampling}
	\begin{algorithmic}[1]
        \State $i \gets -1$
        \State $t_{\text{curr}} \gets 0$
        \State $\mathcal{R}_{\text{sample}} \gets \{\}$
        \ForAll{$t \in \mathcal{T}_{switch}$}
            \If{$t \geq T_{\text{curr}}$}
                \ForAll{$n \in [1, 1 + \lfloor\frac{t-t_{\text{curr}}}{T_{\text{sample}}}\rfloor]$}
                    \State $\mathcal{R}_{\text{sample}} \gets \mathcal{R}_{\text{sample}} \cup \{\mathcal{R}[\max{(0, i)}]\}$
                    \State $t_{\text{curr}} \gets t_{\text{curr}} + T_{\text{sample}}$
                \EndFor
                \State $i \gets i + 1$
                \If{$t_{\text{curr}} > T_{\text{tot}}$}
                    \textbf{break}
                \EndIf
            \Else
                \State $i \gets i + 1$
            \EndIf
        \EndFor
        \If{$ \lfloor\frac{T_{\text{tot}}-t_{\text{curr}}}{T_{\text{sample}}}\rfloor > 0$}
        \ForAll{$n \in [1, \lfloor\frac{T_{\text{tot}}-t_{\text{curr}}}{T_{\text{sample}}}\rfloor]$}
            \State $\mathcal{R}_{\text{sample}} \gets \mathcal{R}_{\text{sample}} \cup \{\mathcal{R}[\max{(0, i)}]\}$
        \EndFor
        \EndIf
    \State{$\textbf{return} \; \mathcal{R}_{\text{sample}}$}
    \end{algorithmic}
\end{algorithm}

\section{Metal Oxide Memristor Modelling}
\label{sec:metal_oxide}
In this section, we look at the application of the modelling framework to the modelling of resistive drift phenomenon under zero bias in titanium dioxide memristors. As such, we give an illustrative example of the proposed modelling framework, detailing a rate equation (dictating switching probabilities), the volatility state variables, and a readout equation for this setting.

\subsection{Physical Model}
\label{sec:rram_model}

Metal oxide \ac{MIM}, and more specifically - Titanium Dioxide \ac{MIM} - has been previously modelled as a single conductive pathway of varying resistance \cite{strukovMissingMemristorFound2008}, and other similar devices, such as tantalum oxide memristors have been modelled as exhibiting the dominance of a single filament or conductive region \cite{strachanStateDynamicsModeling2013}.

Based on initial investigation of a drift dataset (see Section~\ref{sec:dataset} below), we model the conductance of the device as linearly proportional to state variable \(n\).
We hypothesise that the mechanism underlying the switching of the device modulates the width of a conductive region within the active substrate, bridging a connection between the electrodes. Phenomena that could account for this behaviour include a single conductive filament that grows or reduces in width, or multiple conductive filaments that each contribute linearly to the conductance of the device.

\begin{figure}
    \centering
    \includegraphics[width=\linewidth]{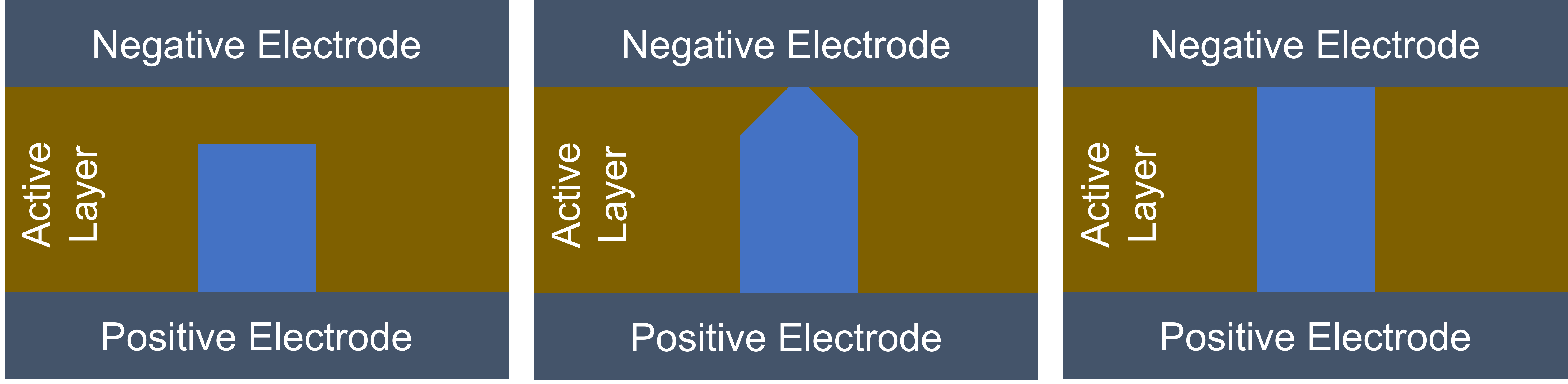}
    \caption{Our physical model for switching in a titanium dioxide memristor. The growth of the filament will not significantly impact the resistance (conductance) - i.e. the readout value - until the filament size crosses a particular threshold. Once this condition is achieved (middle figure), we are in the active region of switching, where small changes in the state variable have a significant impact on the resistance, due to the modulation of an effective area of conduction through the active layer of the device.}
    \label{fig:physical_model_titanium}
\end{figure}

Our model suggests that a large number of switches must be in the low resistance state before there is any measurable impact on the rate. Switching a sufficient number of the switches into the low resistance state can be thought to be performed during the electroforming step. We can imagine our model being a conductive filament that grows in length, eventually making physical contact with the electrode on the other side of the active layer, allowing electrical conduction to occur. Once this contact occurs, additional switching causes the contact region to grow linearly in area of contact, resulting in a linear increase in the conductive state once we are in this ``active'' region. The model for this process is visualised in  Figure~\ref{fig:physical_model_titanium}.

\subsection{Readout Equation}

In the case of Titanium Dioxide devices, we assume that the state variable constitutes the number of parallel filaments that have been formed through the device, or alternatively, the diameter of a single conductive filament. In both cases, the resistance will not be linearly related to this state variable. Rather, it is the conductance, thus our readout equation will be:

\begin{align}
    R(n) = \frac{1}{g_{\text{diff}}\max{(n - n_{\text{thresh}}, 0)} + g_{\text{parallel}}}
    \label{eq:resistance_readout}
\end{align}

We can see from the equation that the behaviour of the model is thresholded, with the conductance remaining at the value \(g_{\text{parallel}}\) until the value of \(n\) surpasses the parameter \(0 \leq n_{\text{thresh}} \leq N\), upon which the conductance increases linearly in steps of \(g_{\text{diff}}\). This is motivated by the conductive filament model described above.

We can also invert the equation to find \(R\) in terms of \(n\):

\begin{align}
    n(R) = n_{\text{thresh}} + \frac{1}{g_{\text{step}}} \left(\frac{1}{R} - g_{\text{parallel}}\right)
    \label{eq:resistance_readout_inverted}
\end{align}

This equation is useful for determining the correspondence between a certain resistance value and the number of switches in the ``low'' state, \(n\), for example, for determining the desired \(n\) value for a specified equilibrium resistance. Note that if the resistance is at its maximum, we simply assume the value of \(n\) to be \(n_{\text{thresh}}\), although it could be any value at or lower than this. This is based on the assumption that we lie in or close to the active switching region once the device has been electroformed.

\begin{figure}
    \centering
    \includegraphics[width=\linewidth]{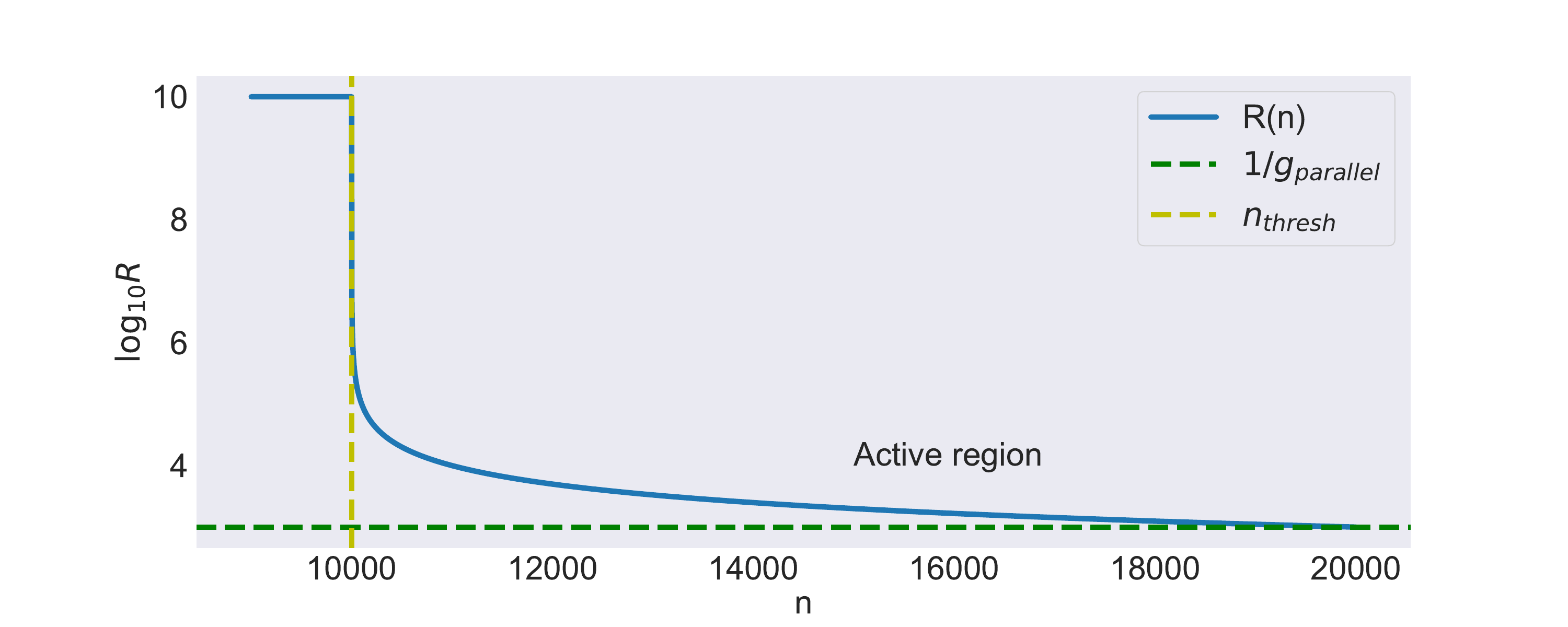}
    \caption{The active switching region is defined and shown as the region where changes to the state variable measurably impact the readout variable (in this case the resistance). Before this point, changes to the state are modelled not to have an impact.}
    \label{fig:active_switching}
\end{figure}

Note that though the readout equation is not linear in the state variable, the equivalent conductance (reciprocal of the resistance) is linear in the state variable, the physical motivation for which is provided above. From the readout equation, we can see that the active region will begin at approximately \(n=n_{\text{thresh}}\). This is visualised in Figure~\ref{fig:active_switching}.

\subsection{Switching Probability}
\label{sec:switching_probability}

To determine the rates of the Poisson processes, we make use of a Boltzmann occupation function that is dependent on the voltage bias and the temperature, as well as static parameters, rather than the Fermi-Dirac occupation function (logistic function) used to determine the switching probabilities in the \ac{GMSM}. This is due to it being a good model of energy-barrier modulated processes at thermodynamic equilibrium in classical systems. %

Equation~\ref{eq:individual_switching_rate} below, gives the Poisson rate of a single switch transitioning from the ``high'' - or ``open circuit'' - state, \(m\), to the ``low'' - or ``conducting'' - state, \(n\): \(\lambda'_{\text{down}}\). Note that in the equations, two physical constants are also used: Boltzmann's constant, \(k_B\), and the electronic charge, \(q\). As a result, Equation~\ref{eq:switching_rate} gives the rate parameter for the overall up switching process.
\begin{align}
    \lambda'_{\text{up}}(t) &= \exp{\left(-\frac{E_{a}-V(t)q/2 - E_{\text{off}}/2}{k_B T(t) (1+\rho(t))}\right)},
    \label{eq:individual_switching_rate} \\
    \lambda_{\text{up}}(t) &= n(t)\lambda'_{\text{up}}(t),
    \label{eq:switching_rate}
\end{align}
with the parameters defined in Table~\ref{tab:parameter_definitions}. In this model, application of a voltage bias, \(V(t)\), lowers or raises the activation energy barrier in one direction, simultaneously raising or lowering it in the other direction.

\begin{table}[]
    \centering
    \begin{tabularx}{\linewidth}{|c|X|}
        \hline
        \hline
        \(E_{a} = qV_a\) & Activation energy (barrier height). Increases/decreases the up/down switching rates simultaneously. \\ \hline
        \(E_{\text{off}} = qV_{\text{off}}\) & Energy offset that causes a shift of the equilibrium point from equal switches being in the high/low resistive state. Adjusts the balance of switching rates under zero bias. (Added, rather than subtracted, for \(\lambda'_{down}\)). \\ \hline
        \(V(t)\) & Voltage across the device at time \(t\), which will be added for \(\lambda'_{down}\) as opposed to being subtracted. This gives an additional energy offset when scaled by \(q\). \\ \hline 
        \(T(t)\) & Time-dependent temperature variable in degrees Kelvin. Influenced by the history of applied voltages and the device state history, which directly impact the switching rates, as well as the equilibrium point when \(E_{\text{off}} \neq 0\). \\ \hline
        \(\rho(t)\) & Volatility state variable. Has a similar impact to the temperature in affecting the global switching rate (indeed, temperature is a volatility state variable in its own right). \\
        \hline
    \end{tabularx}
    \caption{Parameters and inputs to the Poisson rate parameter model for titanium dioxide memristive devices.}
    \label{tab:parameter_definitions}
\end{table}

\subsection{Volatility}
\label{sec:metal_oxide_volatility}

In the case of thin film memristors, volatility can be decomposed into two variables. The first of these is the temperature (that may increase due to Joule heating) and the second is a dimensionless volatility state variable to account for additional volatile effects, such as disruptions to the internal structure of the memristor on application of a bias voltage and subsequent restoration of order following its removal (due to interfacial energy minimisation).

In modelling the effects of Joule heating, we can assume that a proportion of the current flowing through the device (proportional to the square of the current flowing through the device) is converted to heat energy and raises the internal temperature. We can also assume that Newton's law of cooling applies simultaneously, resulting in power dissipation and restoration of the internal temperature to the equilibrium (bath) temperature. One can define the rate of change of temperature in terms of a thermal time constant that is a function of a thermal resistance (\(R_{th}\)) and thermal capacitance (\(C_{th}\)) \cite{roldanThermalModelsResistive2021}, dictating how easily heat energy in the form of temperature is captured and dissipated from the device, with the thermal time constant denoted \(\tau_{th} = C_{th }R_{th}\). The differential equation for the internal temperature of the device is given as:

\begin{align}
    \frac{dT}{dt} = \left(T_{\text{bath}} + R_{th}\frac{V^2(t)}{R(n)} - T\right)/\tau_{th}
    \label{eq:rate_of_change_temperature}
\end{align}

where \(R(n)\) is the instantaneous resistance of the memristor (as a function of the state variable, \(n\)).

As an example of a form of additional volatility due to disruption of the structure, we assume that the voltage applied results in a bias-dependent equilibrium volatility, and the volatility approaches this value according to a differential equation, where the rate of change of the volatility is dependent on the difference between the instantaneous volatility and the equilibrium value.
The volatility and its evolution is defined by the following differential equation:
\begin{align}
    \frac{d\rho(t)}{dt} = \left(c_{\text{volatile}} \cdot V(t) - \rho(t)\right) / \tau_{\text{volatile}}
    \label{eq:volatility_diff_eq}
\end{align}

where \(\tau_{\text{volatile}}\) is the volatility time constant, dictating how quickly the equilibrium value is attained (and therefore how slowly a non-zero volatility decays in the absence of an input signal); and \(c_{\text{volatile}}\) is the volatility factor, which - when multiplied by the applied bias - yields the equilibrium volatility (dictating the volatility state variable's leverage on the switching rates). To incorporate the continuous time volatility, we adopt a discrete time sampling approach, with deterministic events at intervals of the given timestep \(t_{\text{sample}}=0.1s\). To compute changes to the parameters, we use Euler integration.

\subsection{Dataset}
\label{sec:dataset}

To evaluate the behaviour of the model under controlled conditions, we created a dataset based on the evolution of memristive states \cite{el-geresyResistiveDriftTitanium2024}.
We collected 2760 device hours of data across two separate memristor arrays. We set each of the devices in the array to a random initial resistance and measured the evolution of the resistance over time, through Kelvin sensing that used a 30ms pulse of 0.3V magnitude to measure the instantaneous resistance of the device, every 10 seconds, over varying numbers of hours.
The devices used were thin-film Titanium Dioxide (\(TiO_2\)) memristors with \(Al_2O_3\) interstitials (the same as \cite{strukovMissingMemristorFound2008}), placed on the diagonal of crossbar arrays to mitigate sneak path currents. The measurements were obtained using the ArCONE memristor characterisation platform \cite{arcinstrumentsltdArCONEMemristor2019}.

Originally, a total of 219 data series were collected. After collection, the series were inspected and a data cleaning step was performed. As part of this step,  series for memristors which had failed or degenerated to a non-volatile state after programming, as well as clearly anomalous measurements within series due to measurement errors, were removed. Following data cleaning, 23 series were excluded from the dataset, leaving 196 series.

\subsubsection{Quantisation}
\label{sec:quantisation}

\begin{figure}
    \centering
    \includegraphics[width=0.8\linewidth]{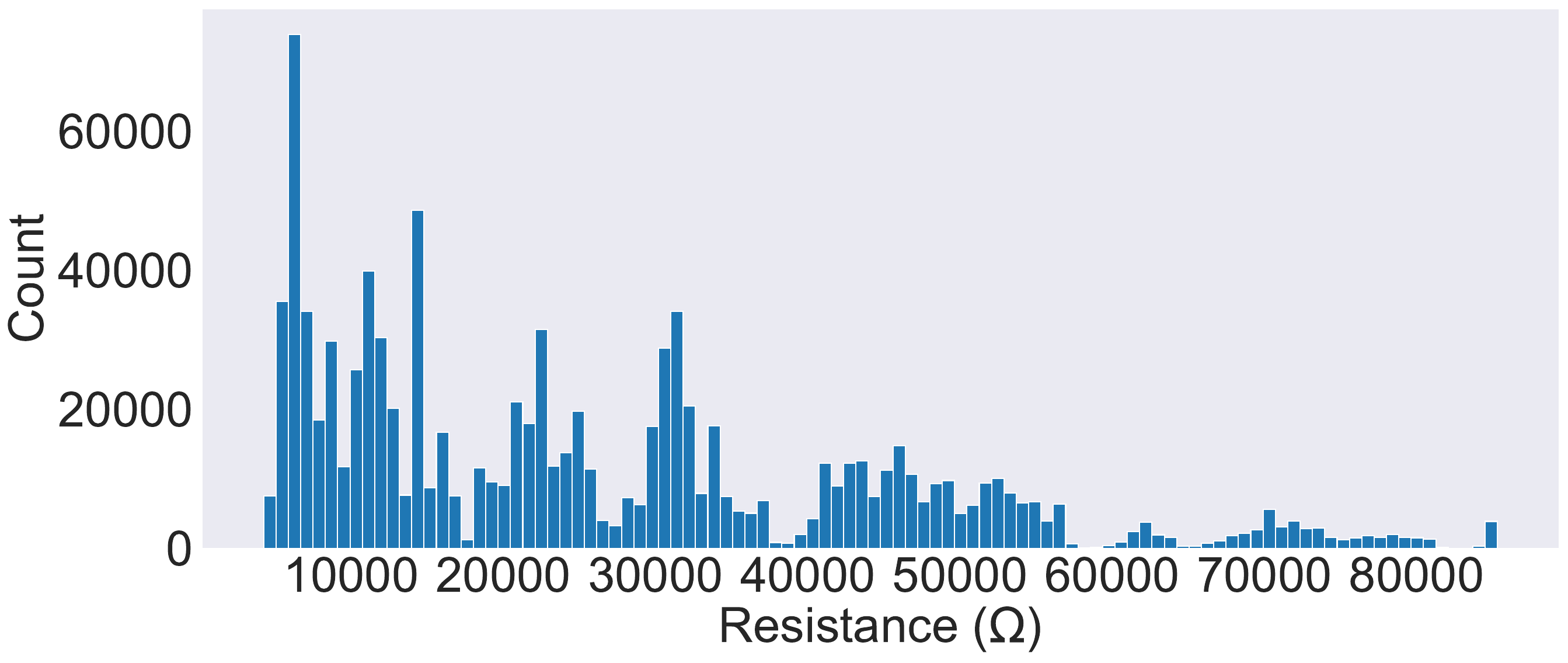}
    \caption{The unquantised resistance dataset, showing the empirical distribution of all resistance values in all series.}
    \label{fig:dataset_histogram_unquantised}
\end{figure}

Figure~\ref{fig:dataset_histogram_unquantised} shows a histogram of the unquantised resistance values in the dataset, showing the range of interest to be the interval of approximately \([10^3, 10^5]\), where changes to the value of \(n\) significantly affect the output resistance. As such, we choose the values of \(N\), \(n_{\text{thresh}}\), \(g_\text{step}\), and \(g_{\text{parallel}}\) accordingly. We set \(g_\text{step} = 10^{-7}\), \(g_{\text{parallel}} = 10^{-10}\), and \(N = 2 n_\text{thresh} = 20,000\).

After choosing the parameter values that determine the number of resistance states and their values, we calculate the resistance corresponding to each possible value of \(n\), using the readout equation given in Equation~\ref{eq:resistance_readout}. Let \(R(n)\) be the readout equation mapping the state variable \(n\) to the corresponding resistance value. The resistance bin boundaries, \(\mathcal{R}_{\text{bound}}\) and values \(\mathcal{R}_{\text{val}}\), in the order of decreasing resistance (increasing state), are given by:

\begin{align*}
    \mathcal{R}_{\text{bound}} = &\{ -\infty, R(n_{\text{thresh}} + 0.5), \\
    &R(n_{\text{thresh}} + 1.5), \cdots, R(N - 0.5), \infty \} \\
    \mathcal{R}_{\text{val}} = &\{ R(n_{\text{thresh}}), R(n_{\text{thresh}} + 1), \cdots, R(N) \} \\
    \text{where}&
    \begin{cases}
        |\mathcal{R}_{\text{bound}}| & = N - n_{\text{thresh}} + 2, \\
        |\mathcal{R}_{\text{val}}| & = N - n_{\text{thresh}} + 1. 
    \end{cases}
\end{align*}

\subsubsection{Pairs}
\label{sec:pairs}

For each series of quantised resistance values in the dataset, we obtain pairs of resistance values from the sequential, non-overlapping intervals of length \(T=10,000\) seconds (2.778 hours). This yields a dataset consisting of pairs of initial and final resistance values for the given interval, allowing us to evaluate the statistics of the resistance drift for such an interval, conditioned on the initial resistance value.

\subsection{Model Fitting}

\subsubsection{Calculation of the Equilibrium Point (Metastable Point)}
\label{sec:equilbrium_point_calculation}
\label{sec:steady_state_analysis}

The change of the memristive state in the absence of a voltage bias is usually referred to as the phenomenon of resistive drift, or volatility.
Given the forward switching rate, we can calculate the equilibrium number of switches in the low resistance state at a given time by equating \(\lambda_{\text{down}}\) to \(\lambda_{\text{up}}\). The expression for this is given in Equation~\ref{eq:switching_ratio}, below. A point of dynamic equilibrium is attained when the rate \(\lambda_{\text{down}} = \lambda_{\text{up}}\); that is, the rate of switching from the high resistance state to the low resistance state is equal to the rate of switching in the opposite direction. In the case that we have a number of states greater than 2, the determination of the equilibrium state is less straightforward and the steady state vector of the associated Markov state transition matrix must be identified.

\begin{align}
    \lambda_{\text{up}} &= \lambda_{\text{down}} \nonumber \\
    n_{\text{eq}} e^{-\frac{E_{a} - V(t)q/2 - E_{\text{off}}/2}{k_b T(t) (1+\rho(t))}}
    &= m_{\text{eq}} e^{-\frac{E_{a} + V(t)q/2 + E_{\text{off}}/2}{k_b T(t) (1+\rho(t))}} \nonumber \\
    n_{\text{eq}} e^{\frac{2(V(t)q/2 + E_{\text{off}}/2)}{k_b T(t)(1+\rho(t))}} &= m_{\text{eq}} \nonumber \\
    n_{\text{eq}} &= \frac{N}{e^{\frac{(V(t)q + E_{\text{off}})}{k_b T(t)(1+\rho(t))}} + 1}
    \label{eq:switching_ratio}
\end{align}

We can see that the equilibrium point, as a fraction of \(N\), is independent of the value of \(E_a\), depending only on \(E_{\text{off}}\) (as well as \(T(t)\), \(V(t)\), and \(\rho(t)\)). This shows that the equilibrium point and the rate of attainment of the equilibrium point can be independently adjusted, using \(E_{\text{off}}\) and \(E_{a}\), respectively.

The equilibrium resistance can be determined from the overall equilibrium state via the readout equation.

When external environmental parameters are functions of time, the model's equilibrium point and the rate parameters associated with the switching point processes will also be functions of time, changing even when there are no underlying state changes due to metastable switching events.

Under a zero- or a piecewise constant- bias setting, the phenomenon known as resistive drift occurs for memristors. This can be shown to be a competition between two Poisson processes with linearly varying rates. There is a well-defined equilibrium point towards which resistive drift occurs, which can be calculated by finding the state for which the rates of the competing switching processes are equal. In the steady state, in a constant-bias setting, the temperature will no longer be time varying. In a zero-bias setting specifically, the temperature will be equal to the ambient temperature. The zero-bias equilibrium point therefore has a simplified expression given as follows:

\begin{align}
    n_{\text{eq}} &= \frac{N}{e^{\frac{qV_{\text{off}}}{k_B T}} + 1 }.
    \label{eq:equilibrium_in_terms_of_voff} 
\end{align}

We can express the offset voltage in terms of \(n_{\text{eq}}\) as:

\begin{align}
    V_{\text{off}} &= \ln{\left(\frac{N-n_{\text{eq}}}{n_{\text{eq}}}\right)}\cdot \frac{k_B T}{q} \equiv \ln{\left(\frac{m_{\text{eq}}}{n_{\text{eq}}}\right)}\cdot V_{T},
    \label{eq:equilibrium_in_terms_of_voff_inverse}
\end{align}
where \(V_{T} = \frac{k_BT}{q}\) is a constant, often referred to as the thermal voltage.

\subsubsection{Linear Conductance Approximation}

\begin{figure}[t]
    \centering
    \includegraphics[width=\linewidth]{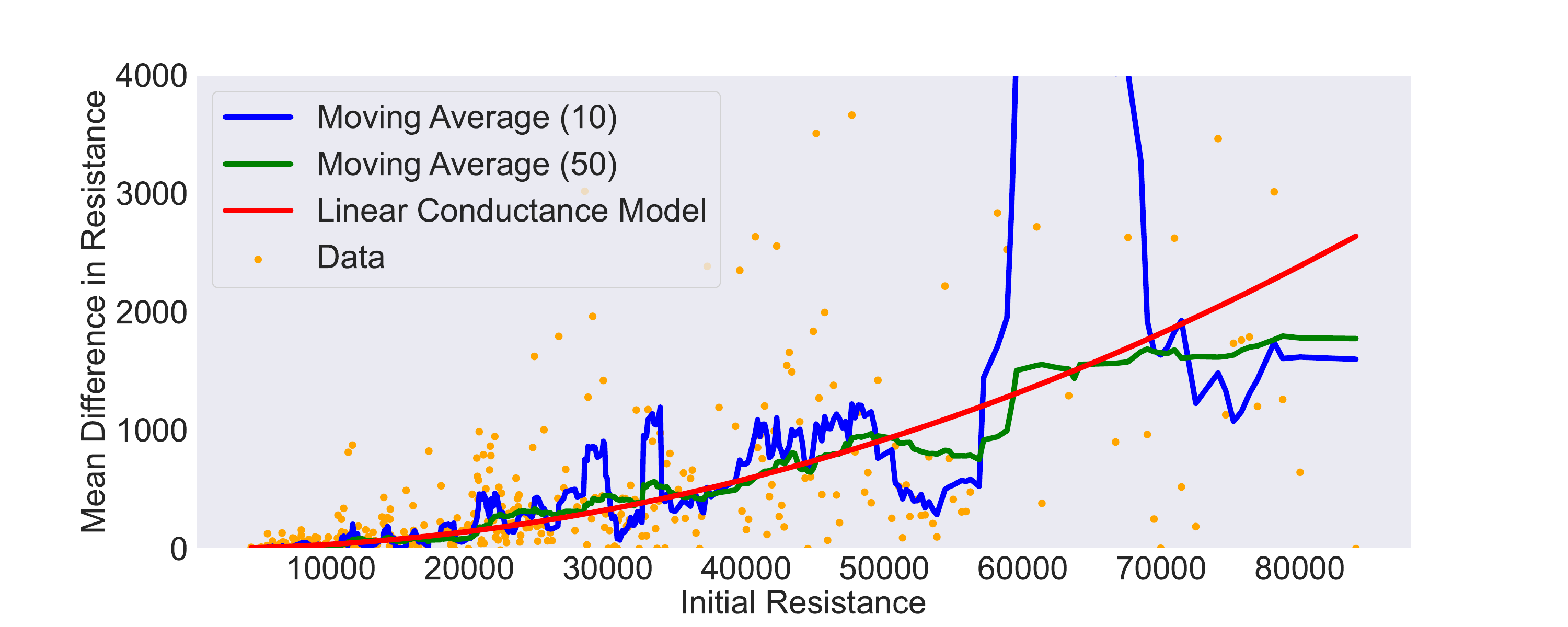}
    \caption{The mean change in resistance for the dataset of pairs with \(\Delta T = 10000\). We show a moving average of the data for different window sizes (5 and 50), to see the non-linear trend.}
    \label{fig:mean_change_in_resistance}
\end{figure}

Based on the physical model described above, we use the approximation that the conductance changes linearly, at a rate proportional to the time. We verify this by measuring the mean change in the conductance conditioned on each of the initial quantised resistance values, \(\mathcal{R}_{\text{val}}\), we calculate the mean change in the resistance across all the pairs in the dataset for given initial resistances. A plot showing this is shown in Figure~\ref{fig:mean_change_in_resistance}.

We use a zero-bias model consisting of an approximately constant change in the conductance over the given time interval, that is independent of the initial resistance. This assumption is based on  a linear difference approximation of our readout equation model (Equation~\ref{eq:resistance_readout}), which fits a trend in the observed data to a reasonable standard. We use an iterative trust region fitting algorithm \cite{branchSubspaceInteriorConjugate1999} to minimise the squared error associated with fitting the function to the predicted resistance change. The model is given as follows, with parameter \(a\) - the change in the state variable - fit to the data:

\begin{align}
    \Delta R(R_{\text{init}}) = \frac{1}{g_{\text{parallel}} - a g_{\text{diff}}} - R_{\text{init}}.
    \label{eq:linear_conductance}
\end{align}

\begin{figure}
    \centering
    \includegraphics[width=0.8\linewidth]{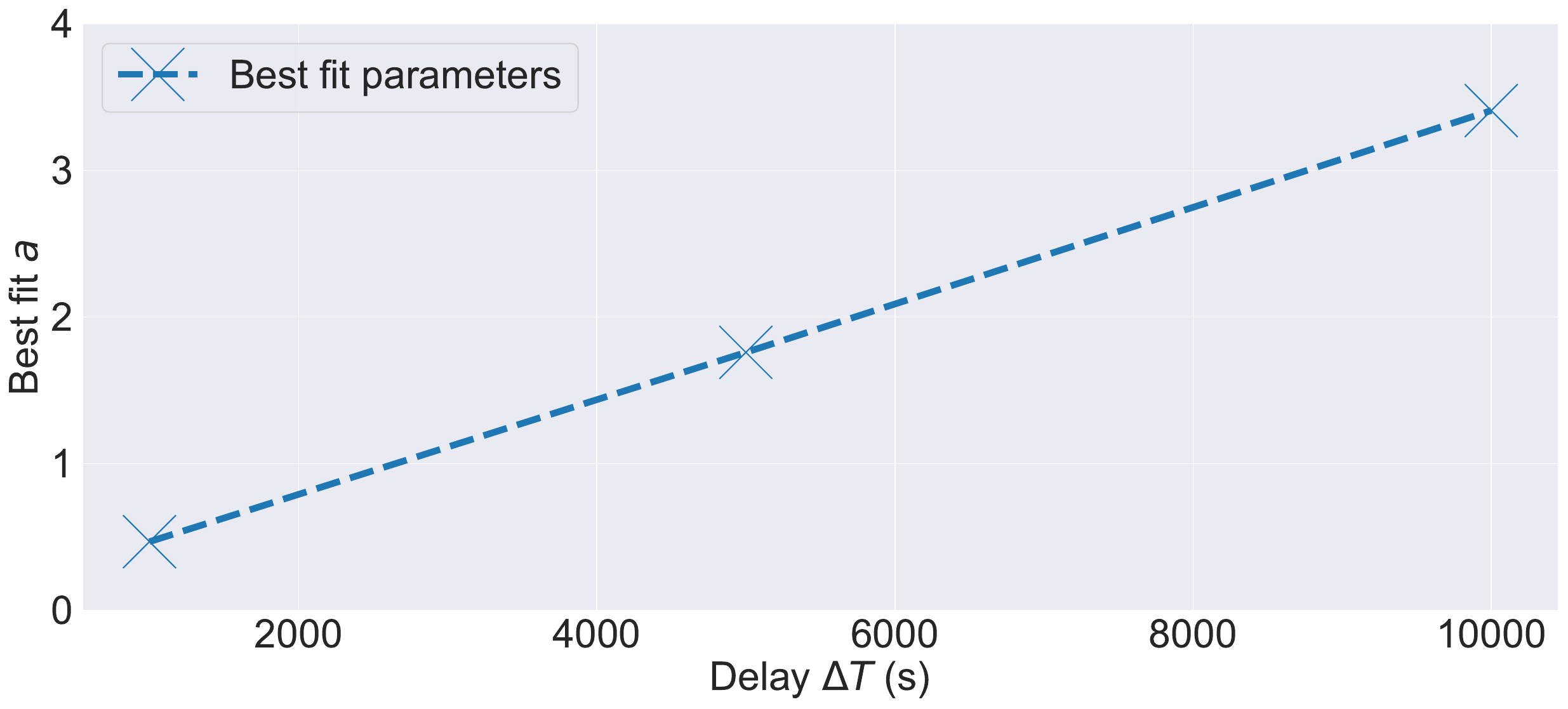}
    \caption{Fitting the parameter \(a\) of the linear conductance model for different timesteps, \(\Delta T\), showing the estimate is proportional to the timestep, supporting the linear assumption.}
    \label{fig:linear_model_timestep_comparison}
\end{figure}

For the chosen timestep of \(10,000\) seconds, we find a value of \(a=3.62\) yields the best fit, for the given parameters \(g_{\text{diff}}\), \(g_{\text{parallel}}\), \(N\), and \(n_{\text{thresh}}\). This associated line of best fit for this linear conductance change model is shown in Figure~\ref{fig:mean_change_in_resistance}. The moving averages are plotted to enable a clearer view of the general trend.
Our linear conductance model is also accurate across different values of \(\Delta T\), as shown in Figure~\ref{fig:linear_model_timestep_comparison}. We choose \(\Delta T=10000\) since larger values of \(\Delta T\) result in datasets with higher signal to noise ratios, since stochastic fluctuations in the memristive state will become smaller compared to the deterministic drift component, allowing for better estimation of the associated model parameter, \(a\).

\subsubsection{Parameter Fitting and Constant Rate Approximation}
\label{sec:parameter_fitting}

We would like to obtain parameters that allow the mean behaviour of the model to correspond to the mean predictions of the linear conductance model.
Due to the difficulty in obtaining an analytic expression from a finite dataset for {the delay-conditional distribution \(P(R(t+\Delta T)|R(t), \Delta T)\) without sufficient knowledge of the nature of the underlying distribution}, we use a constant rate approximation to obtain approximate mean values for this distribution over time. We assume that the up and down switching rates of the model remain relatively constant. In order to fit the model, we calculate the rate at a value of \(n = N\) and make the assumption that the rate does not vary greatly over the region of interest.

As such, we require
\(N >> a\).
We must also ensure that \(N + 1 - n_{\text{thresh}}\), defining the number of unique resistance states according to Equation~\ref{eq:resistance_readout}, is appropriately selected to give a sufficiently fine quantisation over the switching region for which we have data.

\begin{figure}
    \centering
    \includegraphics[width=\linewidth]{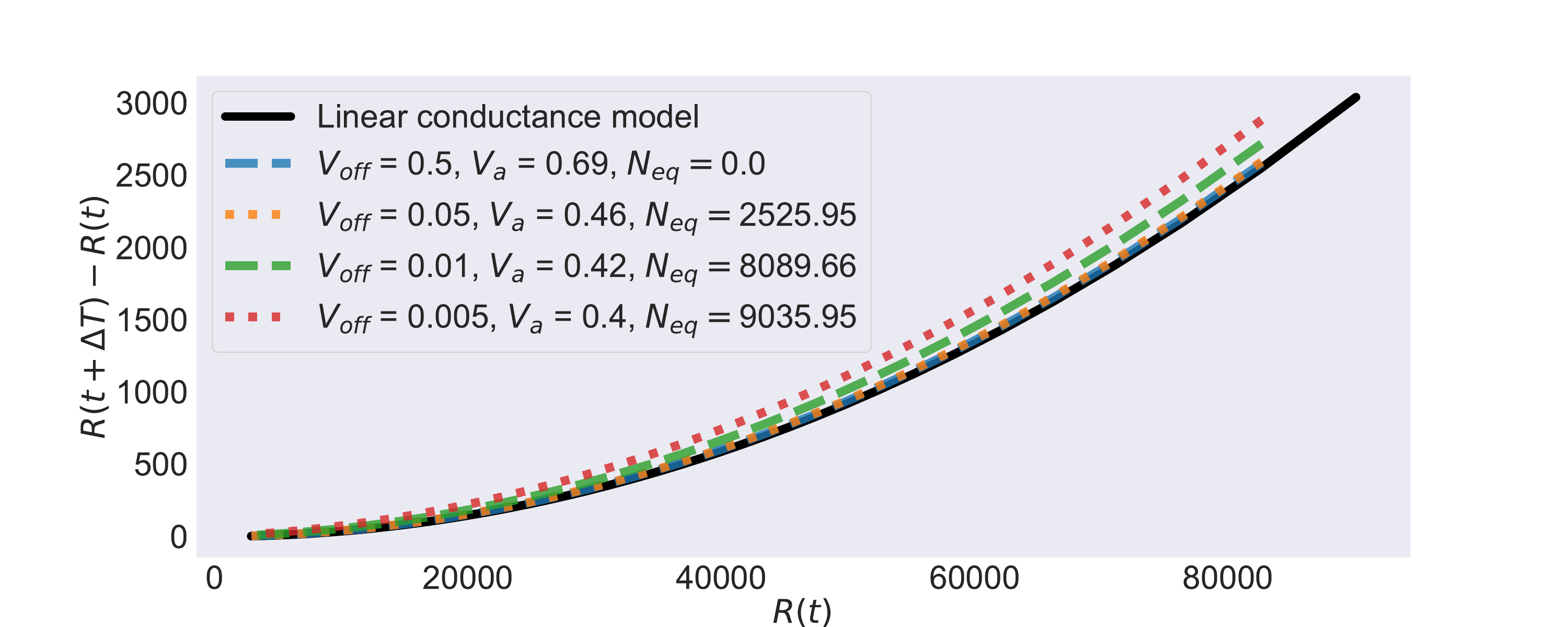}
    \caption{Fitting the event-based model to the linear conductance model. We use a constant rate assumption to derive approximate values for the parameters.}
    \label{fig:fitting_to_linear_conductance}
\end{figure}

For the given value of the parameter \(a\), we can determine an appropriate value of \(V_{a}\). We first select a value of \(V_{\text{off}}\) to ensure an equilibrium point (see Section~\ref{sec:equilbrium_point_calculation}) that is far above the active switching region (in terms of resistance), to ensure that the assumption that the rate is constant over the considered range holds, since the equilibrium point at a particular bias sets the zero-crossing point for the rate equation, hence our constant rate assumption will break down as the value of \(n\) approaches \(n_{\text{eq}}\). In our case, this corresponds to setting an equilibrium point with a value of \(n_{\text{eq}} << n_{\text{thresh}}\). We can see in Figure~\ref{fig:fitting_to_linear_conductance} how the choice of \(V_{\text{off}}\) impacts \(n_{\text{eq}}\); and hence, the constant rate assumption, and the fit to the linear conductance model. Assuming a constant rate and a given value of \(V_{\text{off}}\), we can derive a value of \(V_{a}\) according to the following equation:

\begin{align}
    V_a = \frac{k_B T}{q} \ln{\left( 2 \cdot \Delta T \cdot N \cdot \sinh{\left( V_{\text{off}}\frac{q}{2k_B T} \right)/a} \right)}.
    \label{eq:v_a_constant_rate}
\end{align}

For very small values of \(V_{\text{off}}\), the constant rate assumption will not hold, since the equilibrium point of the model will be high and closer to the active region.

\begin{figure}
    \centering
    \includegraphics[width=\linewidth]{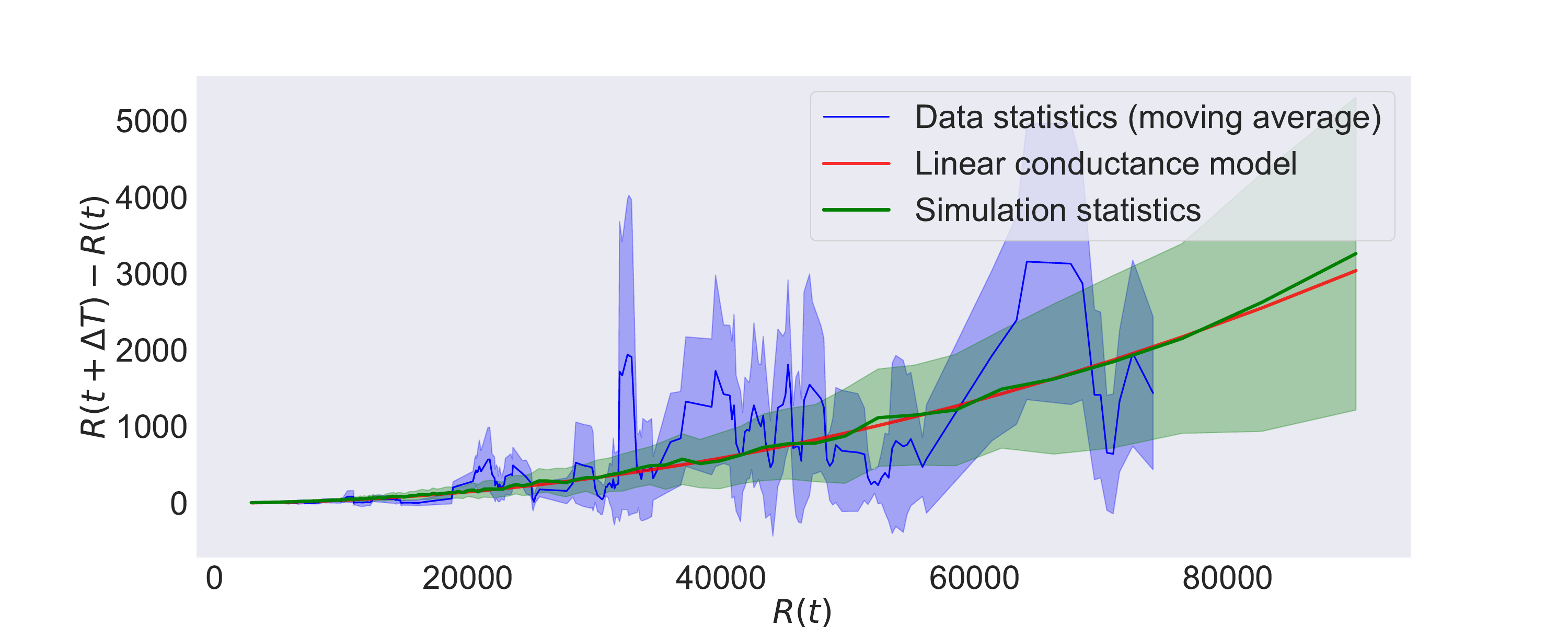}
    \caption{The mean change in the resistance following fitting to the linear conductance model as compared to the data and the linear conductance model, with confidence bands indicating the standard deviation of the data and the simulated series, for simulations with length \(\Delta T = 10000\).}
    \label{fig:model_change_in_resistance}
\end{figure}

In Figure~\ref{fig:model_change_in_resistance}, we see averaged zero-bias simulations from the event-based model, based on the fit parameters, showing the mean change in the resistance fitting to the linear conductance model, as well as the associated variance of the evolutions.

\subsection{Simulation}

In addition to demonstrating the fit of the zero-bias mean values to those determined according to the linear conductance model, we demonstrate the capabilities of the model to simulate memristive behaviour through a series of simulated experiments for a titanium dioxide thin film memristive device. Due to the event-driven nature of the state transition model we conduct our simulations in this section and in Section~\ref{sec:neuromorphic} in Python, rather than a circuit simulation language such as SPICE.

\subsubsection{Experimental Parameters}

The parameters used for the experiments, chosen for \(\Delta T = 10000\), are summarised in Table~\ref{tab:parameters}. The choices for these parameters are given in Sections~\ref{sec:quantisation}~and~\ref{sec:parameter_fitting}.

\begin{table}[ht!]
    \centering
    \begin{tabular}{c|c}
        \hline \hline
        Parameter & Value \\
        \hline
        \(T_{\text{bath}}\) & 300 \\
        \(V_a = E_a / q\) & 0.40049 \\
        \(V_{\text{off}} = E_{\text{off}} / q\) & 0.05 \\
        \(N\) & 20,000 \\
        \(g_{\text{parallel}}\) & \(10^{-10}\) \\
        \(g_{\text{step}}\) & \(10^{-10}\) \\
        \(n_{\text{thresh}}\) & 10,000 \\
        \(\tau_{th} = C_{th} \cdot R_{th}\) & \(3.84\times10^{-14} \cdot 4\times10^{4}\)  \\
        \(c_{\text{volatile}}\) & 10
        \\
        \(\tau_{\text{volatile}}\) & 10
    \end{tabular}
    \caption{Table of parameters for the model, for titanium dioxide memristive devices.}
    \label{tab:parameters}
\end{table}

\subsubsection{Switching Experiments}

\begin{figure}
    \centering
    \begin{tabular}{c c}
    
    \includegraphics[width=0.45\linewidth]{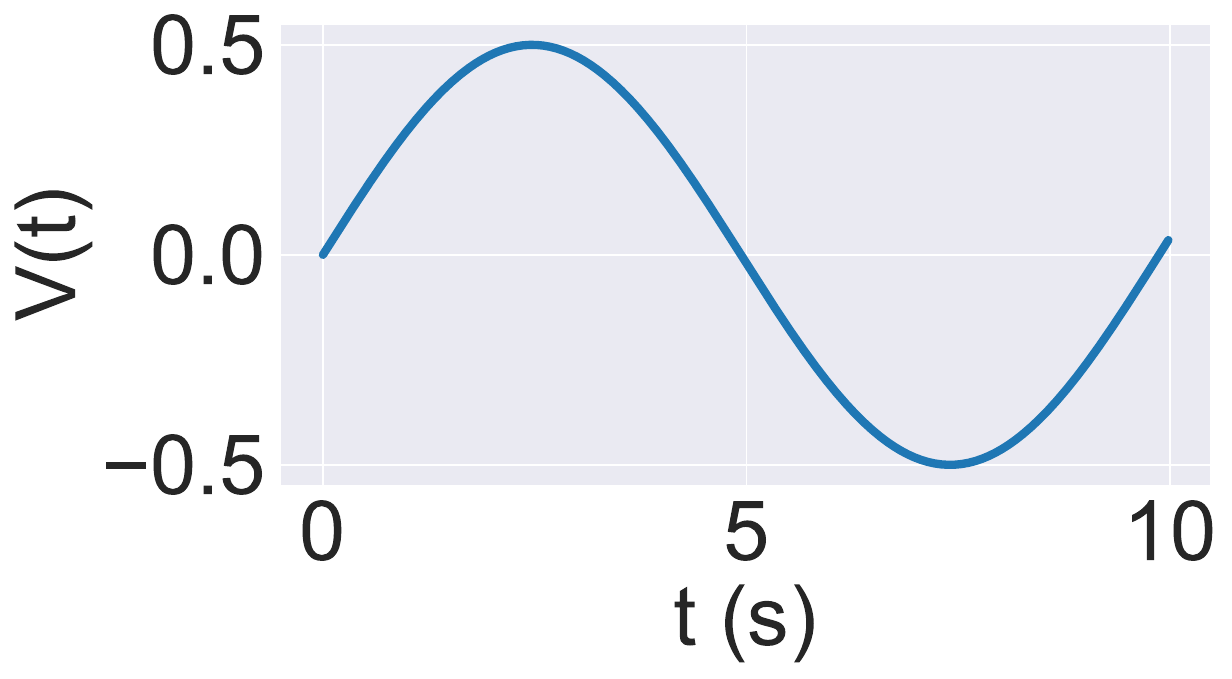} &
    \includegraphics[width=0.45\linewidth]{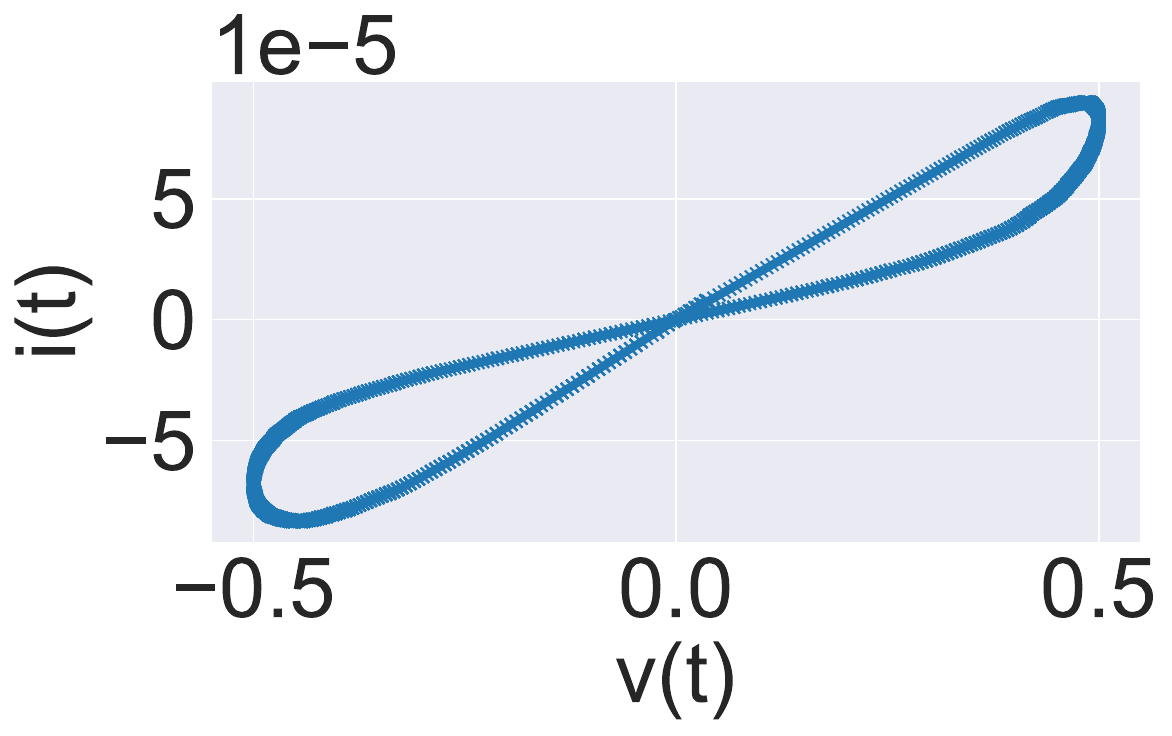}
    \\
    \end{tabular}
    \caption{Memristive hysteresis loop for a device with a starting resistance of 5\(k\Omega\). The input signal is shown on the left.}
    \label{fig:hysteresis}
\end{figure}

\begin{figure*}
    \centering
    \begin{tabular}{c c c c}
     & \textbf{Negative Ascending} & \textbf{Alternating} & \textbf{Positive Ascending} \\
    \includegraphics[width=0.08\linewidth]{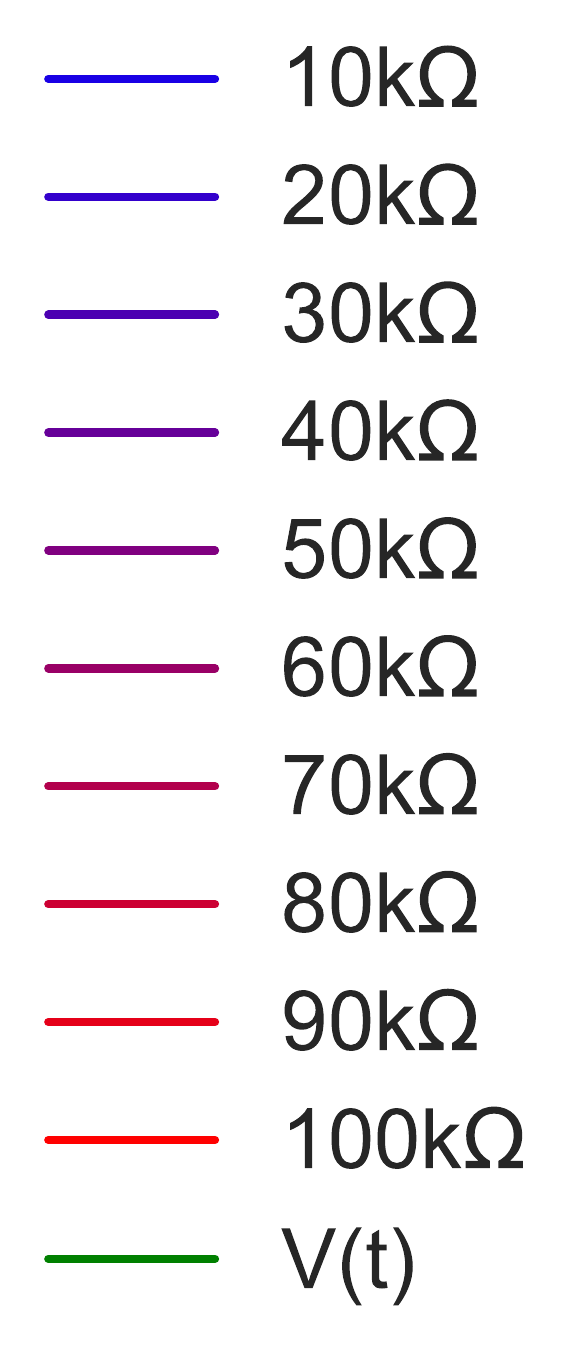} & \includegraphics[width=0.28\linewidth]{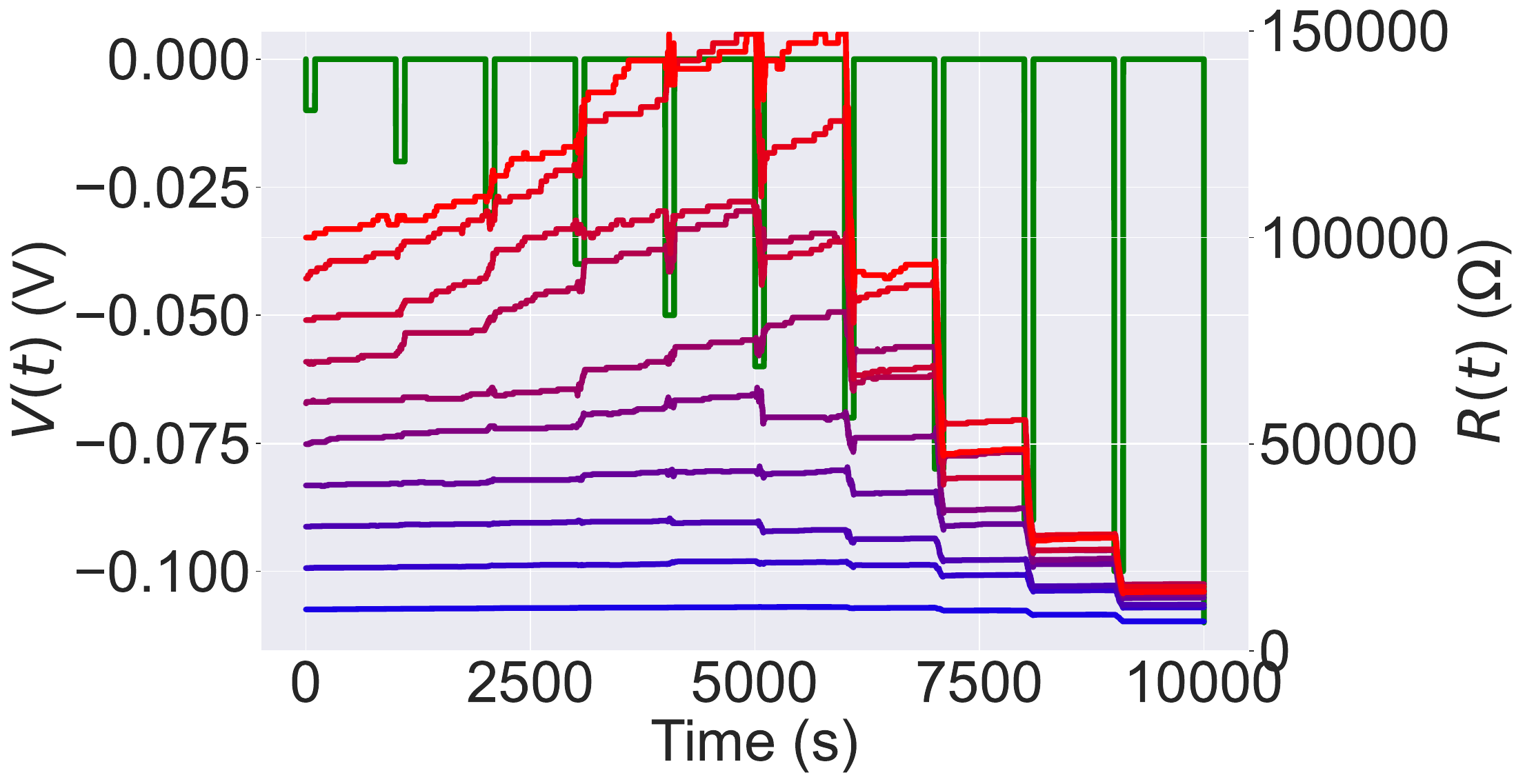} &
    \includegraphics[width=0.28\linewidth]{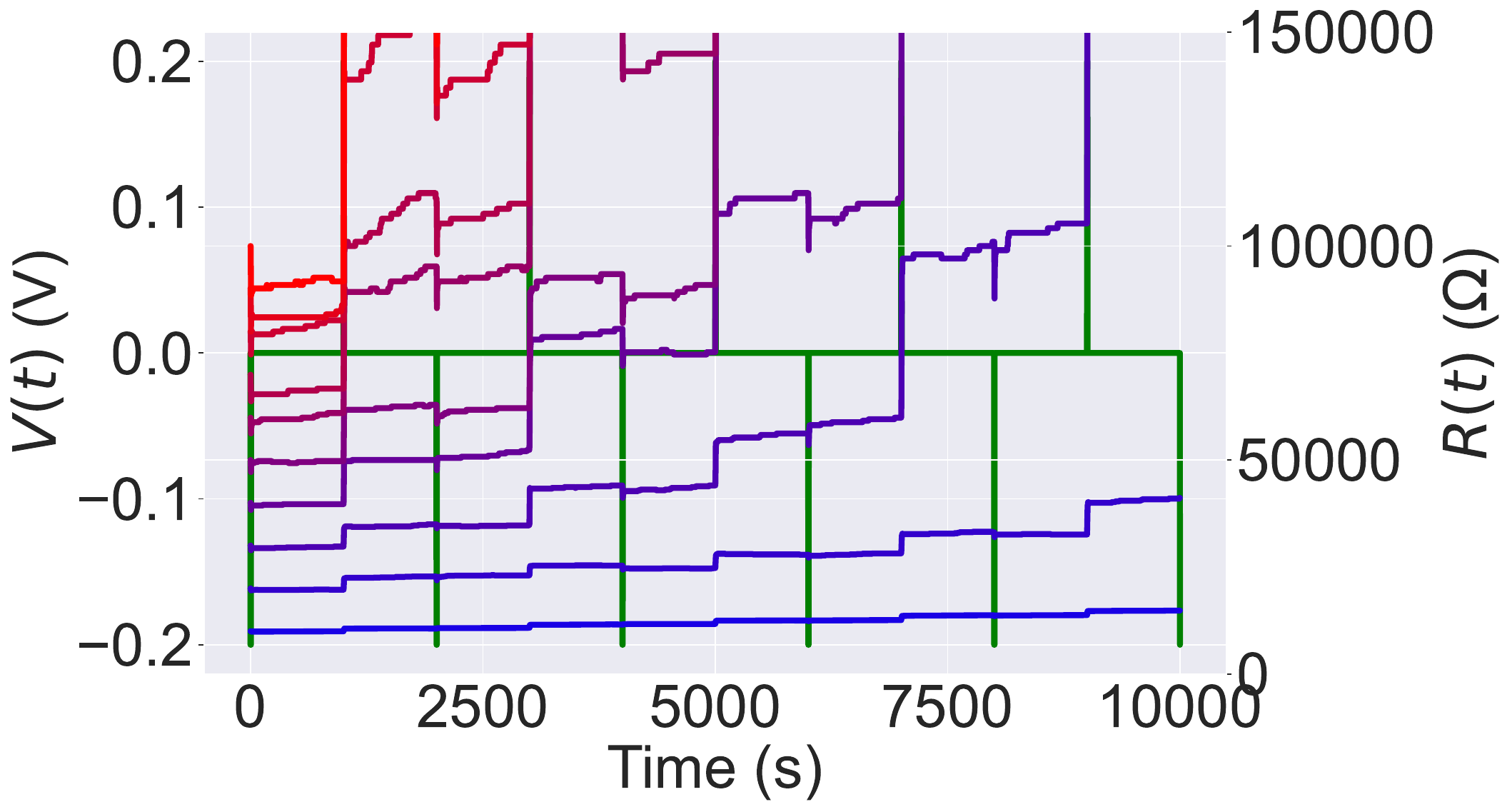} & \includegraphics[width=0.28\linewidth]{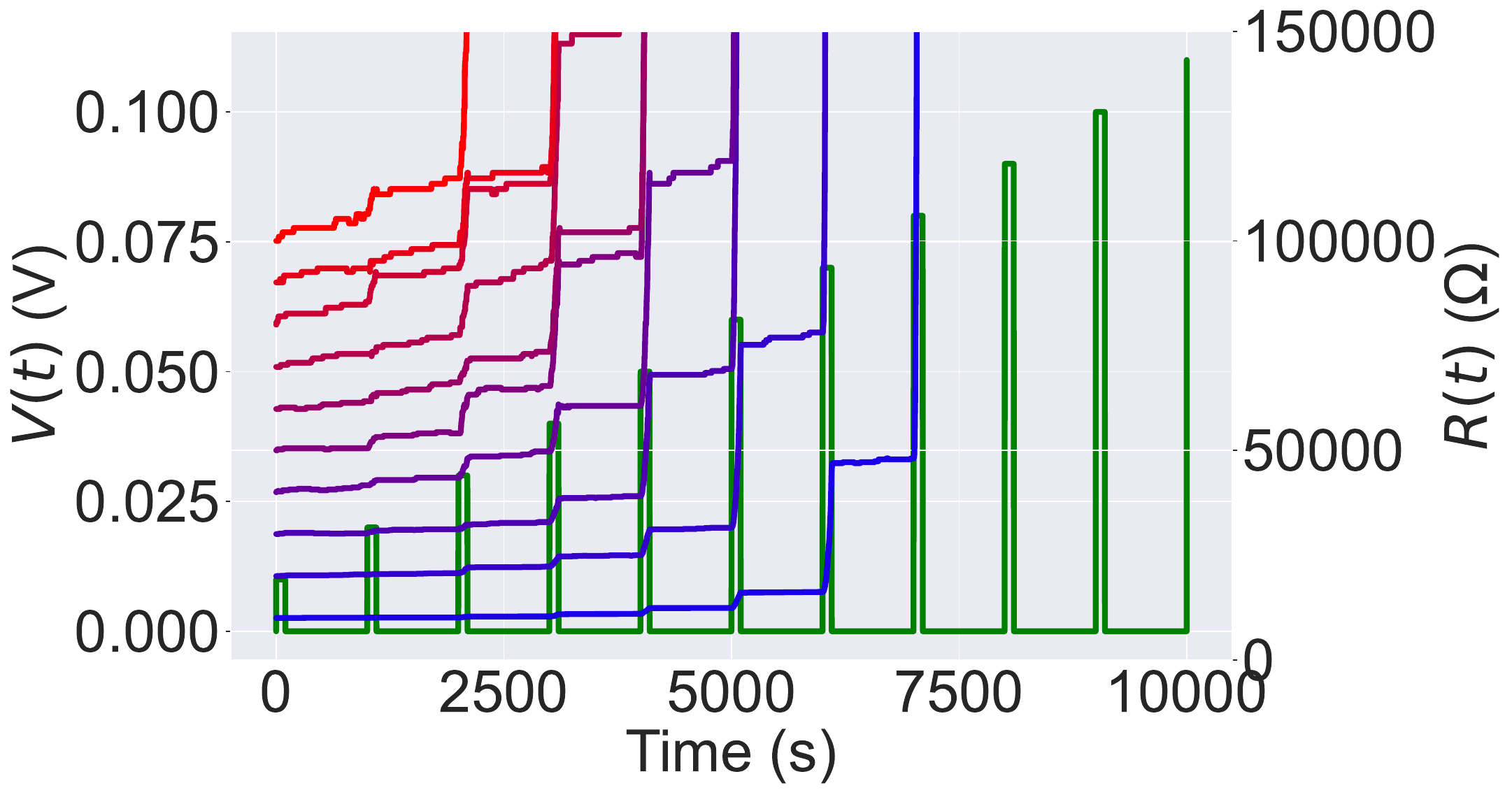}\\
    \end{tabular}
    \caption{Simulated experiments different voltage inputs. We see that the positive/negative pulses switch in the high/low resistance directions, respectively. The high equilibrium resistance results in the final resistance being higher than the initial resistance.}
    \label{fig:switching_experiments}
\end{figure*}

We use the model, with the derived parameters, to generate example device data for simulated input signals.
Figure~\ref{fig:hysteresis} shows a memristive VI hysteresis loop for a device with an initial resistance of 5\(k\Omega\), when a sine wave is applied. In addition to this, we perform three sets of experiments: pulsing with positive voltage pulses (increasing/decreasing in magnitude) to increase the device resistance, and with negative voltage pulses (increasing/decreasing in magnitude) to decrease the device resistance. We also examine the effect of alternating pulses. The results of these switching experiments are shown in Figure~\ref{fig:switching_experiments}.
We simulate series from starting resistances ranging between \(10k\Omega\) and \(100k\Omega\). We choose the parameters for the input signals to highlight model behaviour. In all the experiments, we see that the model is biased towards movement to a higher resistance state, with negative pulses having a lesser impact on reducing the resistive state than positive pulses have on increasing the state. During the inactive period of each pulse, we see the device's resistance drift in the upwards direction as a result of the impact of \(V_{\text{off}}\).

\section{Model Accuracy and Efficiency}
\label{sec:accuracy_and_efficiency}
\label{sec:model_discussion}

Here, we evaluate aspects of the accuracy and the computational efficiency of the model, in the context of its applicability to neuromorphic computing.

\subsection{Impact of Parameter $N$}

One strength of a discrete modelling strategy
is that simulations can be conducted at different resolutions, depending on precision requirements, and the scale of interest. For example, if a memristive device is to be used as a binary storage device, where only the states \(1\) and \(0\) are distinct, then a value of \(N=2\) suffices to model the readout value of interest. In contrast, for a simulation where multiple stable states are to be for information storage, a larger value of \(N\) may be required.
As a consequence of the fact that the variance and mean of a Poisson process are equal to the rate of the process, the magnitude of the parameter \(N\) will scale the expected value and the variance of the final state in a linear fashion.
We can retain the variance and mean by adjusting the parameter \(V_{\text{a}}\) to offset the impact of \(N\) on the simulation statistics. In the case of aligning our model to the phenomenon of drift, as previously shown in Figure~\ref{fig:model_change_in_resistance}, we can use Equation~\ref{eq:v_a_constant_rate} to set \(V_{\text{a}}\) to keep the rate parameter the same, for a given \(V_{\text{off}}\) and updated value of \(N\).
In Figure~\ref{fig:simulation_n_comparison}, we compare resistive drift simulations for different values of \(N\), with \(V_a\) adjusted to achieve the same mean and variance (rate parameter). Runtimes vs. errors for the simulations are shown in Figure~\ref{fig:discrete_time_errors}.

\subsection{Discrete-Time Sampling Comparison}

We note that the approximation in Equation~\ref{eq:molter_model} is valid only for small values of the discrete timestep \(h\). If \(h\) grows large, then discretisation errors emerge. The larger \(h\) becomes, the more significant a change will have occurred in the parameters used to calculate the transition probabilities, leading to inaccuracies.
An advantage of our event-based adaptation of the \ac{GMSM} is to eliminate errors associated with discrete time-stepped simulation.
However, it should be noted that using a value of N that is too coarse will still introduce a quantisation error.

\begin{figure*}[htbp]
    \centering
    \begin{subfigure}[b]{0.32\textwidth}
        \includegraphics[width=\textwidth]{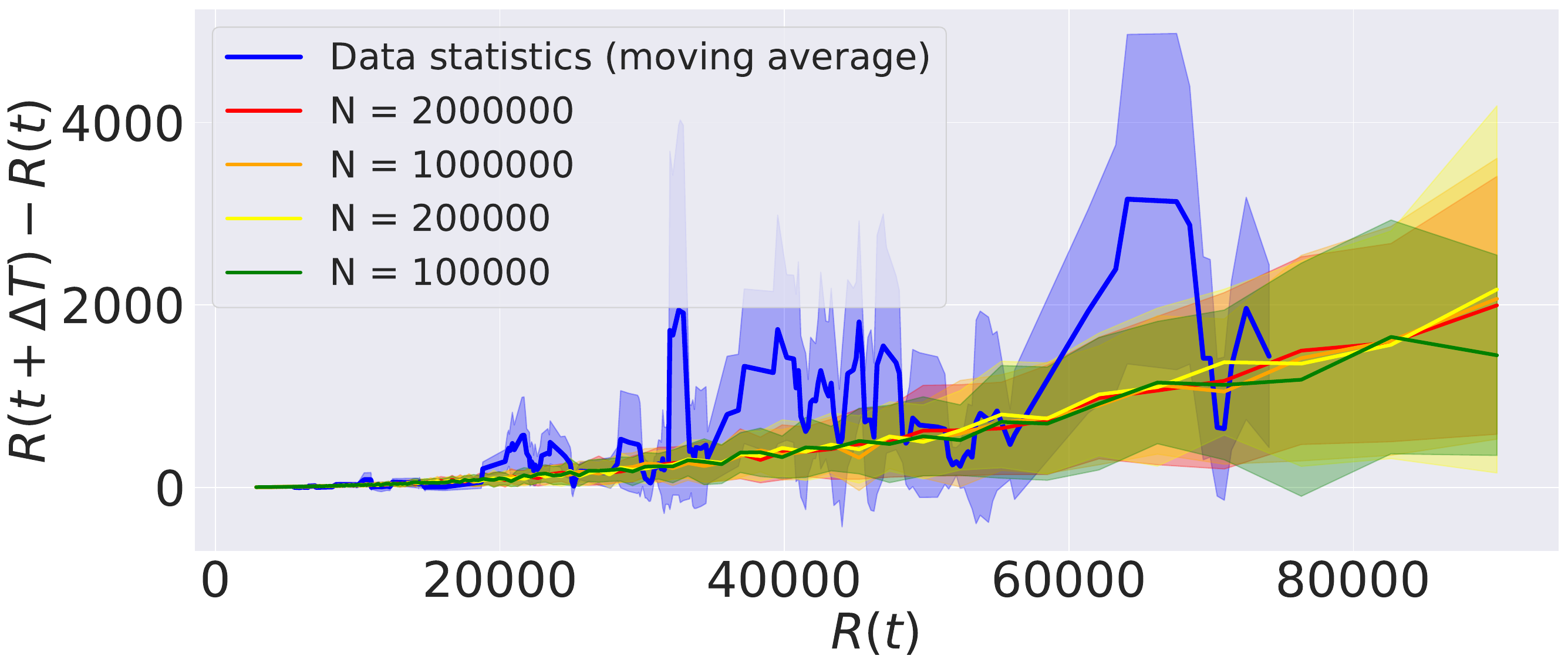}
        \caption{Simulations for different resolutions (\(N\) values), with \(V_{\text{a}}\) adjusted to give equivalence.}
        \label{fig:simulation_n_comparison}
    \end{subfigure}
    \hfill
    \begin{subfigure}[b]{0.32\textwidth}
        \includegraphics[width=\textwidth]{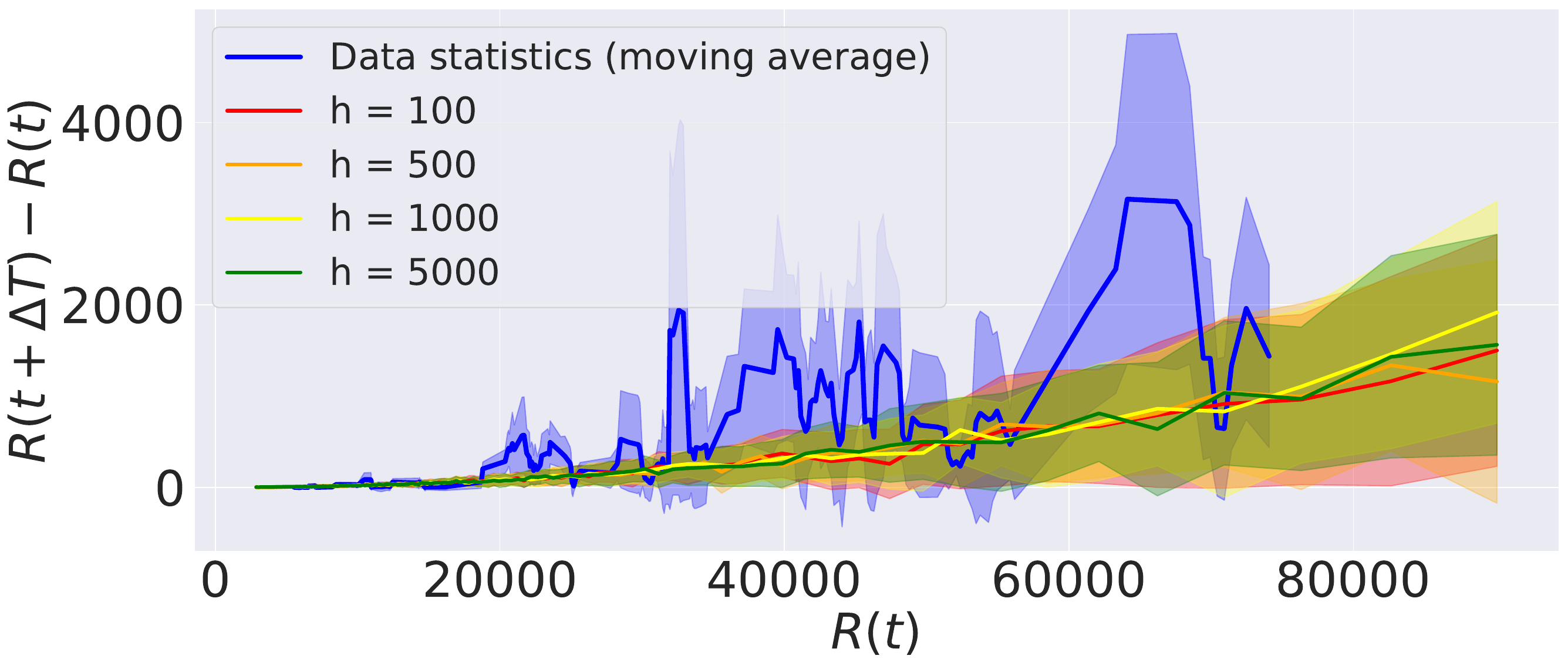}
        \caption{Discrete time approximations of the event-based simulation, for increasing \(h\).}
        \label{fig:discrete_time_simuations}
    \end{subfigure}
    \hfill
    \begin{subfigure}[b]{0.32\textwidth}
        \includegraphics[width=\textwidth]{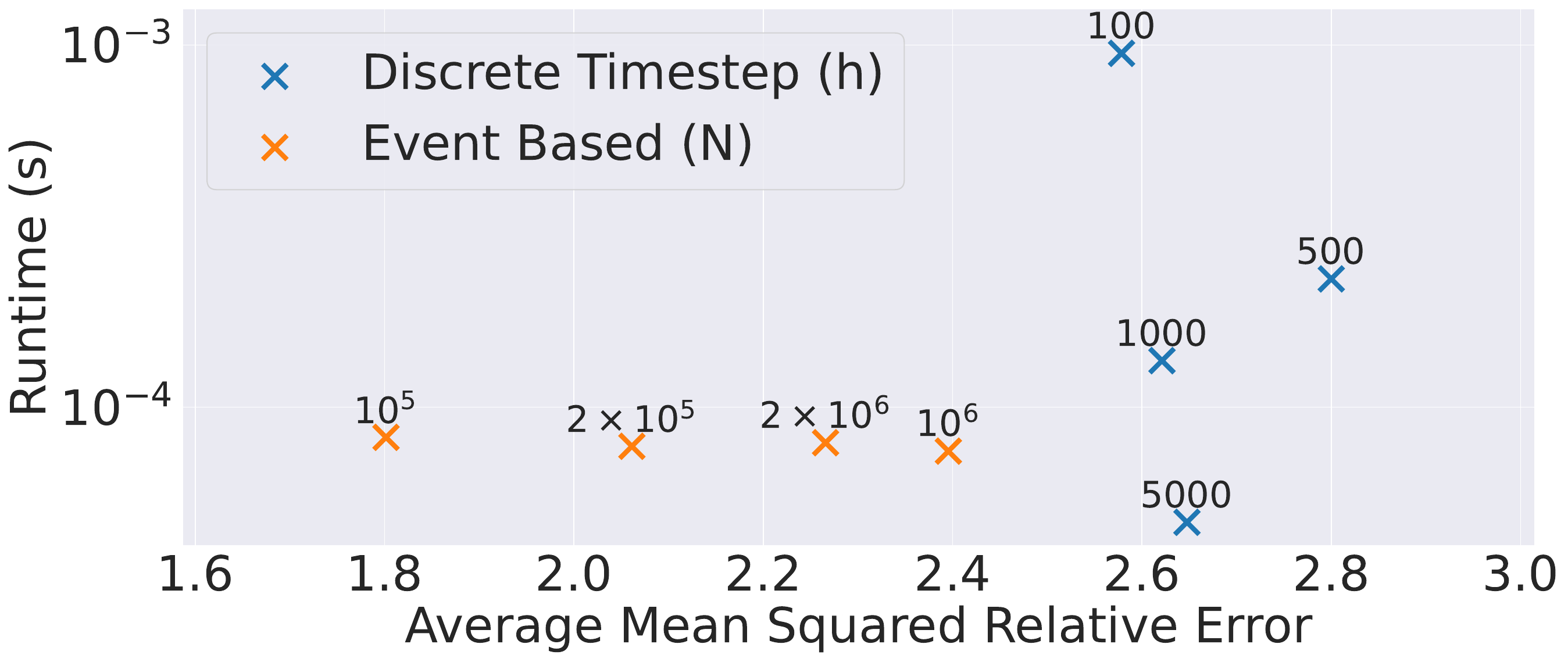}
        \caption{Runtimes vs. average errors for the discrete and event-based modelling approaches.}
        \label{fig:discrete_time_errors}
    \end{subfigure}
    \hfill
    \caption{Comparison of a) event-based for different \(N\), and b) discrete timestep based simulation for different \(h\), with runtime vs. error shown in c). In c), each event-based simulation has the same runtime due to the Poisson rates being the same. We compute the error as the sum of the relative square errors in the mean and variance averaged over initial resistances.}
    \label{fig:main}
\end{figure*}

A discrete time Euler approximation of the event-based switching model can be achieved by using a normal approximation to the number of events generated by a Poisson process over a given timestep \(h\) as \(\mathcal{N}(\lambda h, \lambda h)\).
Figures~\ref{fig:discrete_time_simuations}~and~\ref{fig:discrete_time_errors} show 100 simulations per timestep of the discrete time switching approximation and a corresponding error-runtime curve, demonstrating the trade-off between decreasing the timestep size for accuracy and increasing the computational burden. We see that a discrete time approximation corresponds to the event-based switching to a reasonable degree, and this approach may offer benefits in settings where the number of switching events per unit time is high.
In our case, we limit our modelling to the resistive drift setting, for which the number of switching events is low, leading to the event-based model outperforming the discrete time model in both accuracy and runtime. In the case of incorporation of event-based approximations of continuous parameter updates, such as those for volatility state variables, as outlined in Section~\ref{sec:non_event_based_inputs}, we can upper bound the error of the event-based model with a discrete time modelling error for a timestep \(h\). This frames the event-based modelling approach as an extension of discrete time modelling, with additional parameter updates occurring dynamically where they are prompted by model changes.

Finally, we note that circuit simulation software, such as SPICE, is also capable of implementing adaptive timesteps for input signals, which may offer a good tradeoff between accuracy and speed. However, for simulations involving stochastic state evolution, such approaches are less suitable, as they rely on an approximation of the instantaneous rate of change of parameters to compute the required timestep size, which is not straightforward to estimate for stochastic dynamics.

\section{Neuromorphic Applications}
\label{sec:neuromorphic}
There are a number of ways that the volatility state variable in the modelling framework developed may be used to shed light on approaches that aim to replicate the behaviour of biological neurons using memristors in neuromorphic computing settings. Here, we use the titanium dioxide model realisation we developed in the previous section to elucidate this.

\subsection{Stable and Unstable Memristors}

We may consider two types of memristor: those which are highly \textbf{unstable} and those that are \textbf{stable}. For the model as presented in Section~\ref{sec:metal_oxide}, the stability of the memristor can be characterised by the rate of attainment of its equilibrium point, which is determined in large part by the magnitude of the energy barrier \(E_a\). Unstable devices may be useful for the emulation of neuronal body dynamics, where the interplay between volatility and the application of a stimulating input (for example, current spikes) can result in self-limiting spiking responses. In this context, resistive drift happening over short timescales is analogous to the leakage of membrane potential. In contrast, stable memristors may be more useful for the emulation of synaptic weights. In this context, we can model a combination of short and long term memory effects: \ac{STP} can arise for the programming of a stable memristor to a particular state while the volatility is high. The volatility may cause the state variable to partially revert to its equilibrium point, but eventually dissipates, resulting in stability of the resistive state (\ac{LTP}). For stable memristors, resistive drift is analogous to the phenomenon of synaptic weight decay.

\subsection{Volatility and its Impact}
The inclusion of volatility in a memristive model adds a key characteristic that could prove crucial to the emulation of neuronal dynamics such as metastable leakage terms and spiking behaviour (short-term memory), in addition to more stable and persistent changes to the device state (long-term memory). The volatility has an impact on both the \textbf{equilibrium point} and the rate of attainment of the equilibrium point (the \textbf{time constant of decay}). The impact of the volatility state variable on the equilibrium point for a titanium dioxide model was shown in Section~\ref{sec:equilbrium_point_calculation}.
For stable memristors, the interplay between changes to both rates of switching in the same direction, caused by volatility, and (biased) changes in opposite direction caused by the application of voltage pulses of a given magnitude and time, can lead to memory effects, where states exhibit a dependence on the relative timing of the pulses. A sequence of identical voltage pulses can result in a very different final metastable resistance depending on their spacing in time. For example, in the context of Joule heating, if the voltage pulses are short enough for adiabatic cooling (or relaxation) to occur, the volatility may not increase sufficiently to result in significant switching behaviour. If, however, voltage pulses are more closely spaced, this can result in a larger attained volatility during their application, resulting in switching. Thus, changes in the memristive state are not only dependent on the absolute energy supplied, but also on the way in which this energy is supplied through the pulse shapes and durations.

\subsection{Neuromorphic Simulations}

We here perform simulations demonstrating the impact of the volatility state variable on two classes of memristor. We first demonstrate the possibility of frequency-dependent switching for stable memristors, and then demonstrate the generation of action potentials for unstable memristors.
We choose to use the same rate equation (Boltzmann) and readout equation (linear conductance) modelling components proposed for the TiO2 memristor in Section~\ref{sec:metal_oxide}, and use the second volatility component proposed in Section~\ref{sec:metal_oxide_volatility}: structural disruption, mediated by either the voltage for frequency-dependent potentiation (Section~\ref{sec:frequency_dependent_potentiation}), or the current for spiking behaviour (Section~\ref{sec:spiking_behaviour}).

\subsubsection{Frequency-Dependent Potentiation}
\label{sec:frequency_dependent_potentiation}
In biological neurons, the generation of action potentials can be dependent on the magnitude of the input, according to the leaky integrate and fire neuron model \cite{hodgkinQuantitativeDescriptionMembrane1952, dayanTheoreticalNeuroscienceComputational2001, gerstnerNeuronalDynamicsSingle2014}, where only an input of a sufficiently high frequency may evoke an action potential - a spike. As a result of the frequency-dependence of postsynaptic neuronal action potential generation and the mechanism of spike timing dependent plasticity (Hebb's rule) \cite{hebbOrganizationBehaviorNeuropsychological1949}, the potentiation of a neuronal synapse may also, consequently, be frequency-dependent. Frequency coding is one of a number of coding schemes used for the communication and processing of information in the brain \cite{kandelPrinciplesNeuralScience2013}.

\begin{figure}
    \centering
    \includegraphics[width=\linewidth]{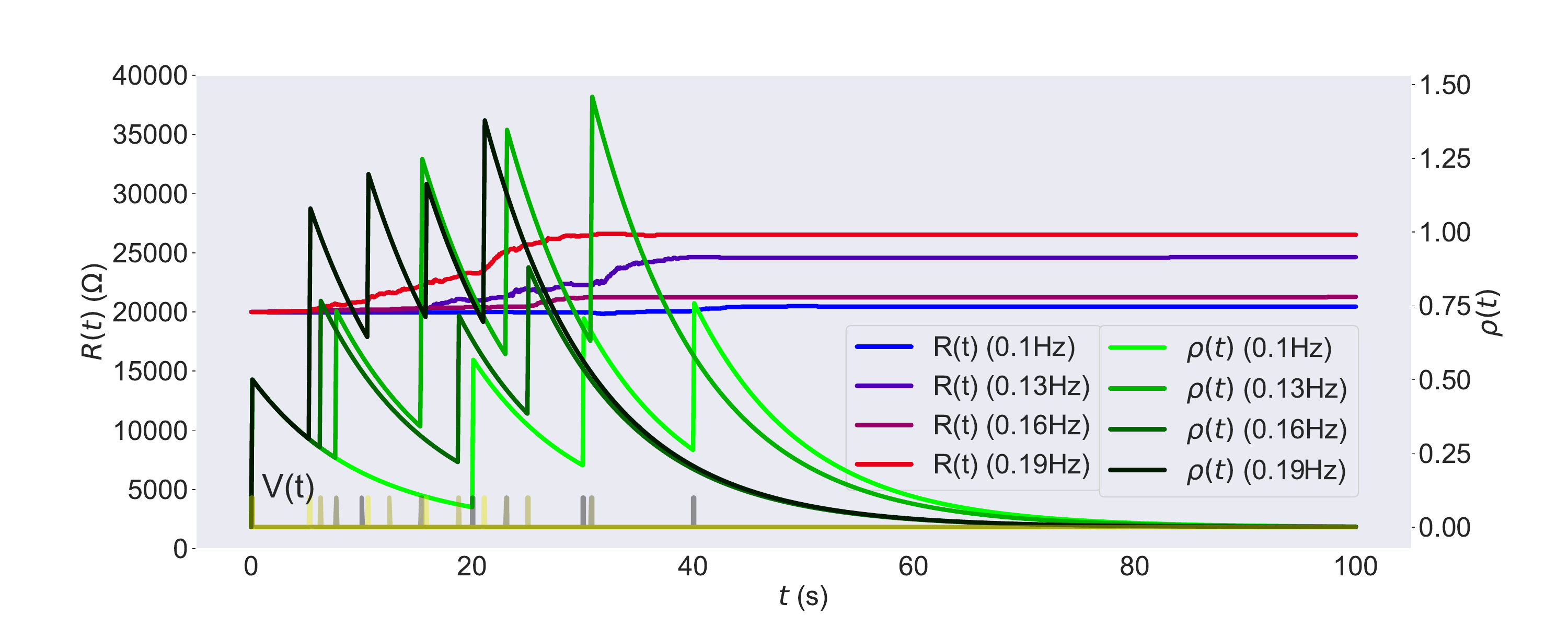}
    \caption{Frequency experiments for a starting state of 20k\(\Omega\). We achieve a higher metastable resistive state when using a higher frequency input for the same number of pulses as a result of the larger magnitude attained by \(\rho(t)\).
    }
    \label{fig:frequency_experiments}
\end{figure}

Frequency-dependent potentiation has been demonstrated in diffusive memristive devices. Diffusive memristors operate on the basis of the dispersal and regrouping of metal ion clusters under voltage bias \cite{wangMemristorsDiffusiveDynamics2017}. Frequency of application of voltage pulses has been shown to have an effect on the number of pulses required to turn a device ON in protein nanowire catalysed diffusive memristors \cite{fuBioinspiredBiovoltageMemristors2020}.
The number of pulses (of equal magnitude and duration) required to switch the devices from high to low resistive states was significantly reduced when the frequency of application was increased from 50Hz to 900Hz. A close to linear trend in the number of pulses required as a function of frequency was observed for frequencies higher than 200Hz, with increasing frequency reducing the number of total pulses required to initiate switching. However, a non-linear trend was observed for sub-100Hz frequencies, where the number of pulses increased rapidly. This could be modelled by a volatility state variable whose time constant of decay was on the order of \(1/100Hz=0.01s\).
In the context of our event-based model, a device with a moderately stable switching characteristic (whose switching rate declines rapidly following the removal of applied bias) can exhibit such frequency dependent behaviour. This is dictated by the state the device is in, as well as the values \(V_{a}\) and \(V_{\text{off}}\).
In Figure~\ref{fig:frequency_experiments}, we show how changes to the frequency of programming pulses can result in different degrees of potentiation in memristors due to the aforementioned interplay between the volatility and switching. For these experiments, we use a simulation duration of \(T=100s\). We incorporate discrete timestep events to the volatility state variable (\(\rho(t)\)), at intervals of \(t_{\text{sample}}=0.1s\), with Euler integration used to perform the updates at events. In these experiments, we supply the same amount of energy to the devices (the same magnitude of voltage over the same amount of time in total), but demonstrate that if the impact of the volatility state variable is high enough, subsequent state transitions can differ greatly, dependent on the frequency of programming, due to the non-linear acceleration in switching effected by an increase in the volatility state variable when the pulses are closely spaced, as compared to sparse pulses in time. We apply square waves of period \(T_{\text{pulse}} = 1/f\), where \(f\) is the frequency of the applied signal. Each square wave period contains a pulse of length 0.1s, and we supply 5 periods as an input signal in each experiment. As such, the same voltage magnitude is applied for the same total amount of time in each case. We modify the parameter \(c_{\text{volatile}} = 500\), keeping all other parameters as in Table~\ref{tab:parameters}.

\begin{figure}
    \centering
    \includegraphics[width=\linewidth]{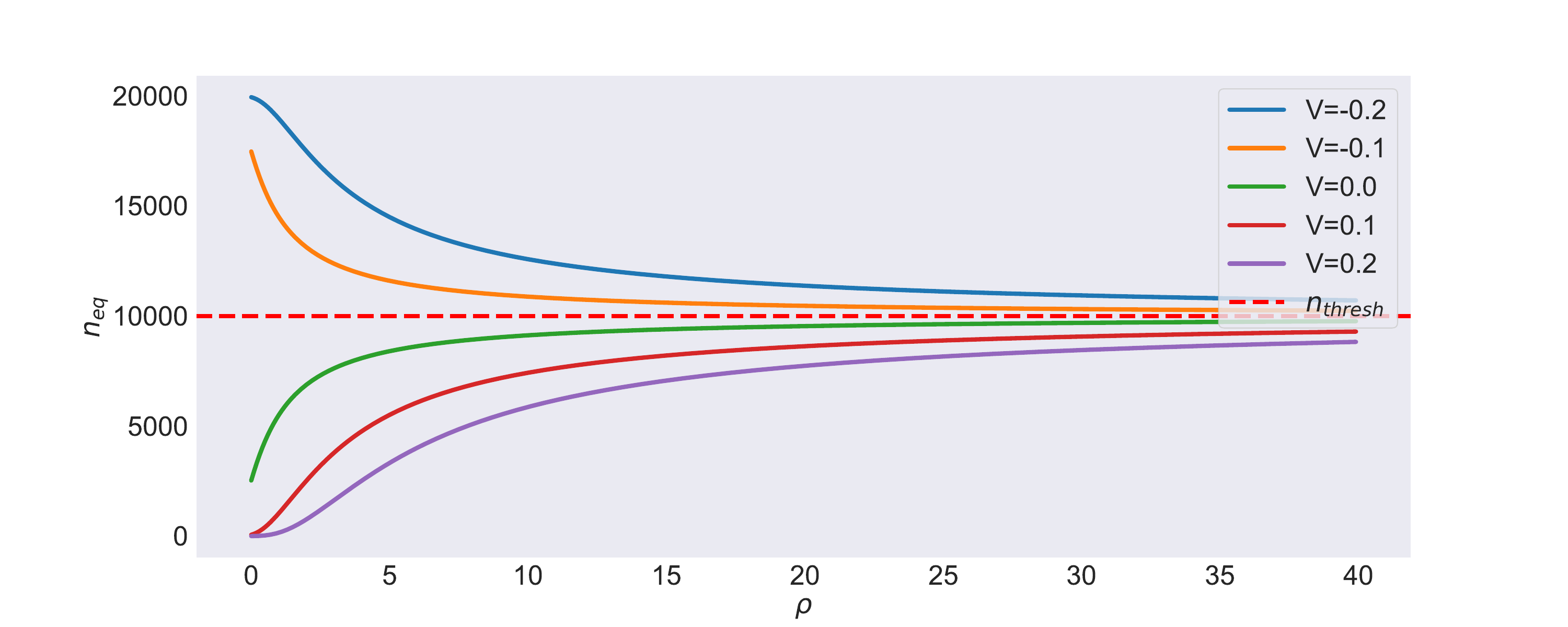}
    \caption{The impact of \(\rho(t)\) on \(n_{eq}\), for different bias values. Parameters for the calculation are drawn from Table~\ref{tab:parameters}. As \(\rho(t)\) increases, the equilibrium point becomes closer to \(N/2\).}
    \label{fig:rho_impact}
\end{figure}

\subsubsection{Spiking Behaviour}
\label{sec:spiking_behaviour}

For unstable memristors, we explore a possible phenomenon enabled by the presence of volatility in a memristive system: the emulation of action potentials. Early on, \cite{chuaMemristiveDevicesSystems1976} looked at the Hodgkin-Huxley neuron model \cite{hodgkinQuantitativeDescriptionMembrane1952} as an example of a memristive system. There has since been interest in action potential generation using memristors \cite{pickettScalableNeuristorBuilt2013}.

A ``spike'' can be defined in the context of a neuromorphic computing architecture as a short-lived increase (or decrease) in an electronic parameter that can be used to process information, such as the voltage, the current, or the conductance. A spike in biological systems is a self-limiting process.
We can envision the creation of spiking behaviour in real devices, if we are able to engineer such a process.
Biological neurons exhibit two responses to constant input stimuli: firing latencies or rates that are proportional to the magnitude of the input signal (time and frequency encoding, respectively), and a firing response that diminishes over time if the input signal remains constant (adaptation). We propose a method for producing such behaviour using the event-based model, through the dynamics of the volatility state variable.

As we can see from Equation~\ref{eq:switching_ratio}, the equilibrium state is a function of the volatility. Regardless of the parameters, an increase in the volatility results in a shift in the equilibrium state towards the point \(n_{eq} = N/2\), as \(\lambda_{\text{up}}\) becomes equal to \(\lambda_{\text{down}}\). Since the magnitude of both the \textit{up} and \textit{down} switching rates goes up, the rate of attainment of the equilibrium point also goes up. This is visualised in Figure~\ref{fig:rho_impact}.

\begin{figure}
    \centering
    \includegraphics[width=\linewidth]{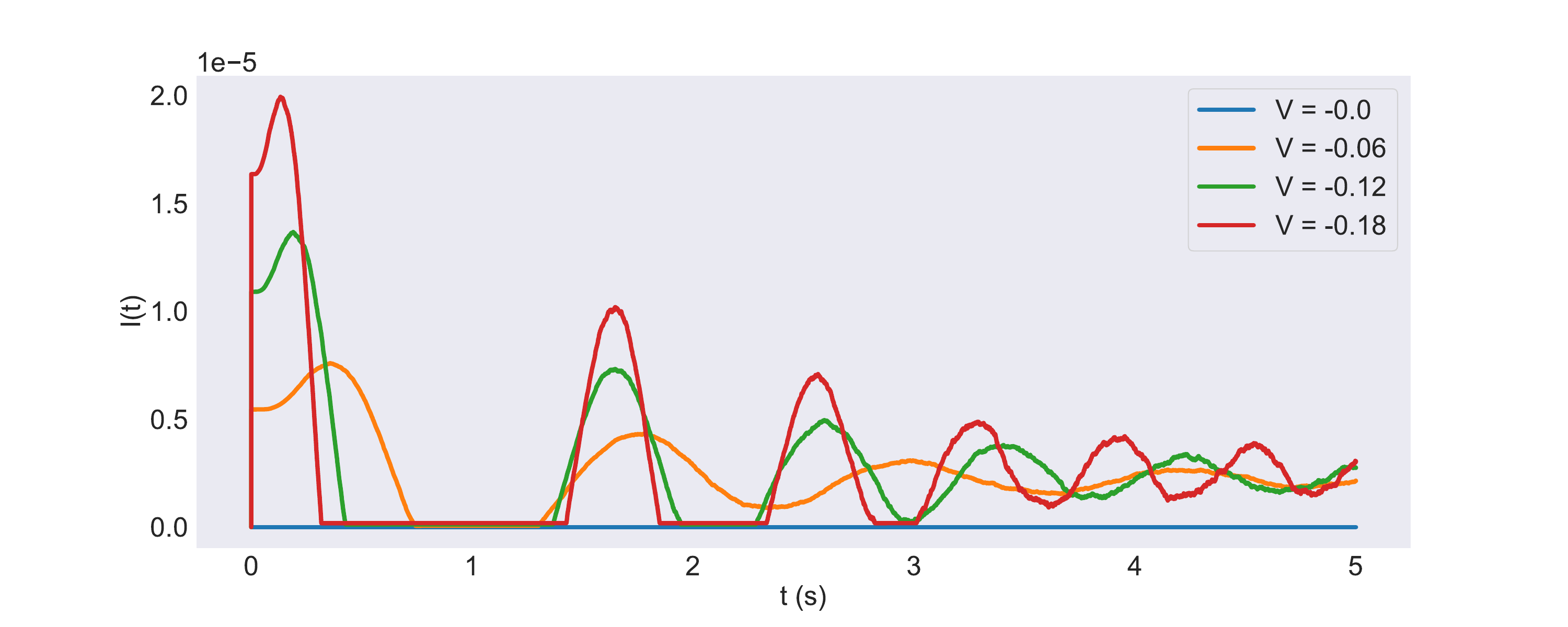}
    \caption{Spiking current behaviour produced using model parameters that produce oscillatory behaviour.
    The controlled oscillations can be mapped to discrete spikes through a thresholding operation. The intensity of the spikes, the number of spikes in the train, and the frequency of the resultant spike train are dependent on the magnitude of the applied input.}
    \label{fig:spiking}
\end{figure}

We can imagine a memristive device with a low resistance equilibrium state at rest. Simultaneously, we can imagine a volatility state variable which increases according to the power dissipated by the device. At low resistances, the power dissipated in the memristor will be high, while at high resistances, the reverse will be true.
We envision the memristor to be in series with a resistor, forming a voltage divider circuit.
At a low resistance, the dissipation of power will result in an increase in the volatility, resulting in an increase in the equilibrium resistance (a drop in \(n_{eq}\)). As the device resistance rises in response, the power dissipated drops. Due to the volatility time constant, the drop in the equilibrium state will not be instantaneous, and the device resistance will continue to rise. Finally, once the volatility decays sufficiently, the equilibrium resistance will approach its original value, and the resistance will decrease, completing a ``spike''. The number of such events that happen per unit time will be related to the magnitude of the applied signal. We model this scenario, using a current-dependent volatility, in Figure~\ref{fig:spiking}.
To produce such a spike train, several parameters were modified, as given in Table~\ref{tab:spiking_modified_params}.
The behaviour is essentially brought about through a controlled oscillatory (self-inhibiting) - or transient - response.
We note the behaviour resembles spiking dynamics in other complex systems that may emerge in the presence of feedback, such as relaxation oscillations in solid state lasers \cite{paschottaRelaxationOscillations2005}.

\begin{table}[]
    \centering
    \begin{tabular}{c|c|c|c|c|c|c}
        \hline
        \hline
        Parameter & \(g_{\text{step}}\) & \(g_{\text{parallel}}\) & \(N\) & \(n_{\text{thresh}}\) & \(V_a\) & \(V_{\text{off}}\)\\
        \hline
         Value & \(10^{-8}\) & \(10^{-6}\) & 1000000 & 800000 & 1.0 & -0.8 \\
    \end{tabular}
    \caption{Parameters to produce spiking behaviour as shown in Figure~\ref{fig:spiking}. Only those differing from Table~\ref{tab:parameters} are listed.}
    \label{tab:spiking_modified_params}
\end{table}

This example demonstrates how through consideration of the volatility state variable as a modelling tool, with its impact on the equilibrium state of the memristor and the rate of attainment of that state, we can predict memory-dependent, self-limiting, short-term behaviour such as spiking, on simulated memristors, offering motivation for its implementation using real memristive devices in a neuromorphic circuit, through the tuning of device parameters. Our perspective offers a principled approach for analysing and engineering this behaviour in memristors.
Though our proposed neuromorphic examples are hypothetical, they offer guidance for the design of future devices through appropriate parameter selection to tune the relationship between the equilibrium point and its rate of attainment, 
helping to decode their behaviour and inform their maturation as devices for neuromorphic computing.

\section{Conclusion}
We have introduced an event-based modelling strategy for memristive devices where time is the main computational resource. In such systems, event-based simulation is essential, and use of a volatility state variable can help to make clear routes to controlling device behaviour through the spacing of input to the memristive device over time, leading to the possibility of the emulation of different forms of neuronal dynamics, dependent on the stability of the memristive device in question.
Our framework, based on multiple parallel Poisson processes, addresses errors associated with discrete time approximation in existing models, as well as being natively suited to event-driven neuromorphic applications.
We proposed a volatility state variable, to allow for the capture of transient switching dynamics, motivating this from a physical perspective for various memristive device types.
We have also presented a resistive drift dataset for resistive drift in titanium dioxide memristors, which we used to obtain parameters for an example of the model applied to a real device, along with proposing a new linear conductance model of drift in the devices as an example of a readout equation, motivated by a model of the connection between the internal state and the observed resistance. We demonstrated the ability of the event-based model to replicate the resistive drift behaviour under zero bias following the development of the Poisson rate parameter model and volatility model for the titanium dioxide memristors, and subsequent parameter fitting.
Finally, we examined the modelling framework from the perspective of neuromorphic computing, studying possible requirements for parameters and control strategies for future devices that could enable frequency-dependent switching, and replication of action potentials using memristive devices.
Future work can include applying the modelling framework described to additional device types, specifying probabilistic switching models, volatility models, and readout equations.

\bibliography{references}

\begin{IEEEbiography}           [{\includegraphics[width=1in,height=1.25in,clip,keepaspectratio]{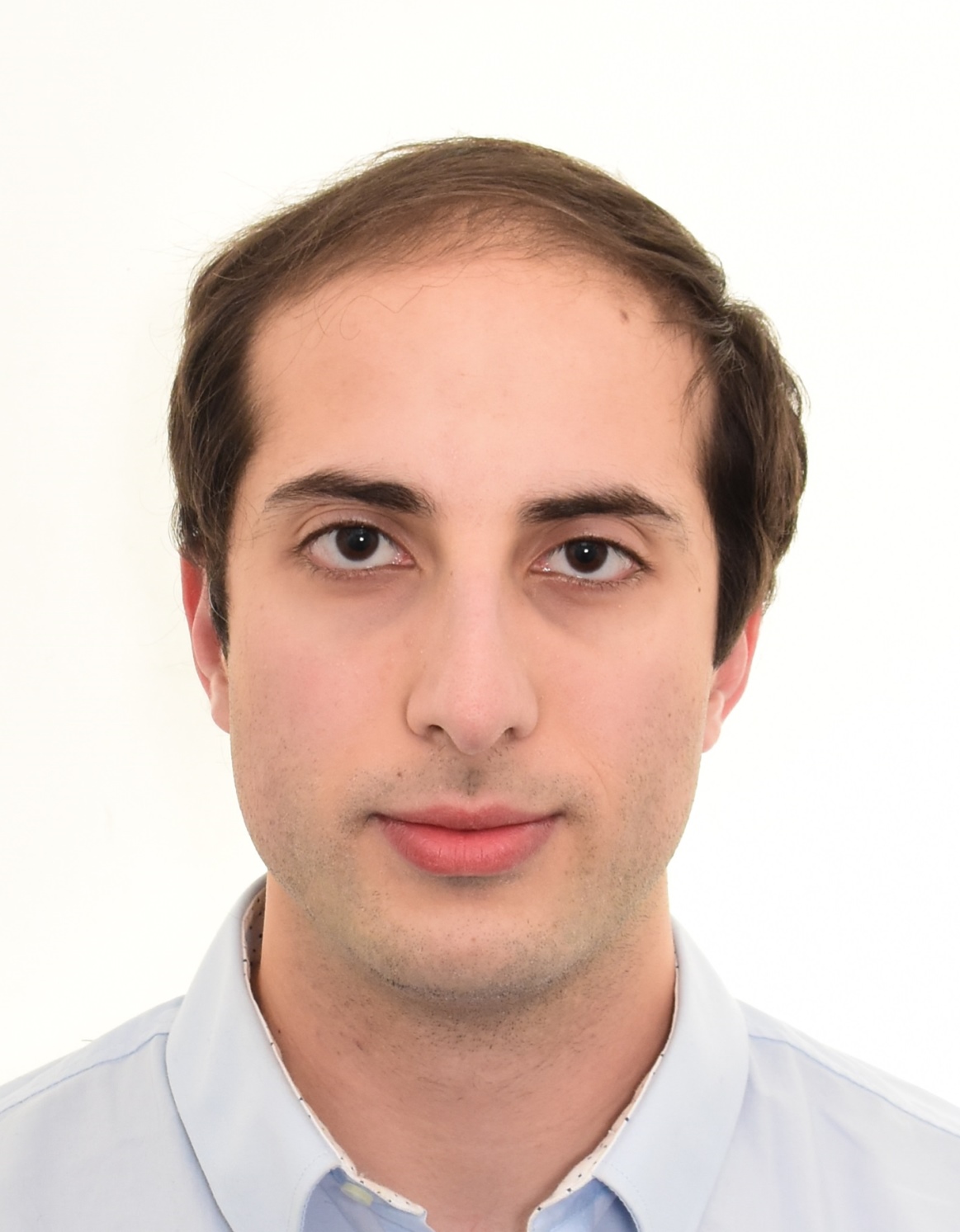}}]
    {Waleed El-Geresy}
    Waleed El-Geresy is a researcher in neuromorphic AI at Nokia Bell Labs, formerly a member of the Intelligent Systems and Networks Group at Imperial College London. He received the MEng and Ph.D. degrees in electrical and electronic engineering from Imperial College London in 2019 and 2025, receiving the best master's thesis prize from the IEEE Information Theory Society. His research interests include information theory, representation learning, neuromorphic computing, memristors, and bio-inspired artificial intelligence.
\end{IEEEbiography}
\begin{IEEEbiography}
    [{\includegraphics[width=1in,height=1.25in,clip,keepaspectratio]{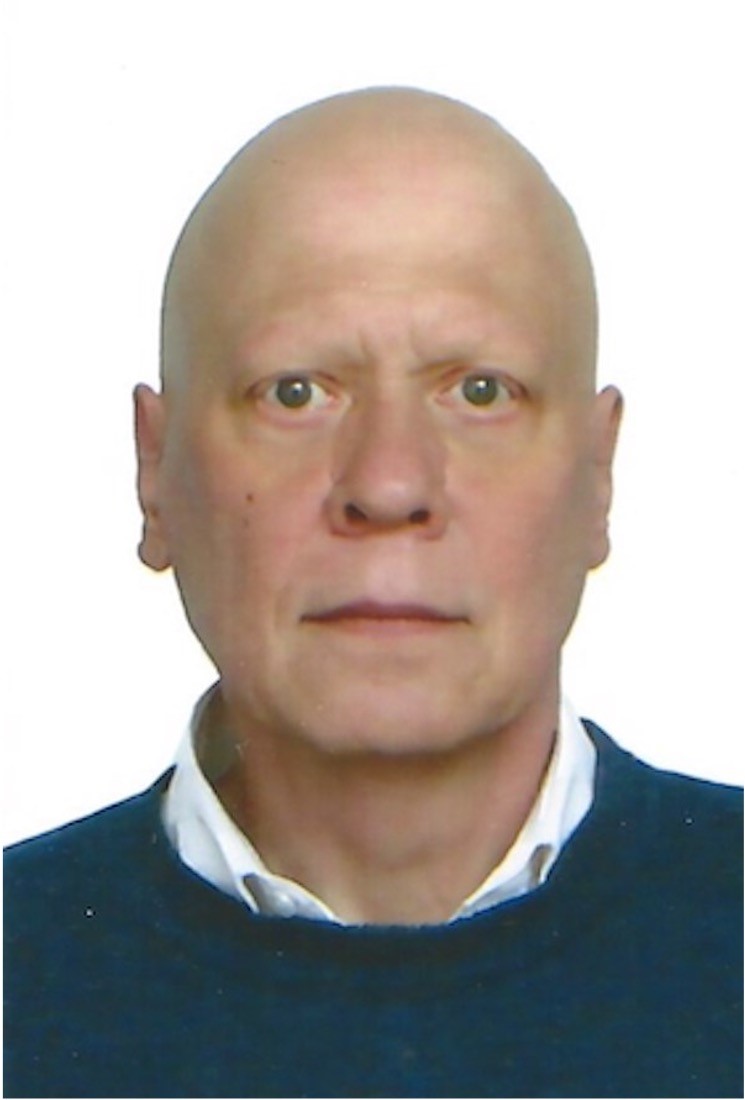}}]
    {Christos Papavassiliou}
    Christos Papavassiliou (M’96–SM’05) is a Reader at the department of Electrical and Electronic Engineering of Imperial College working on Analogue and RF circuit engineering and instrumentation and, more recently, recently on memristor array integration with CMOS and memristor programming and testing interfaces. He received the B.Sc. degree in Physics from the Massachusetts Institute of Technology in 1983, and the Ph.D. degree in Applied Physics from Yale University in 1989.
\end{IEEEbiography}
\begin{IEEEbiography}
[{\includegraphics[width=1in,height=1.25in,clip,keepaspectratio]{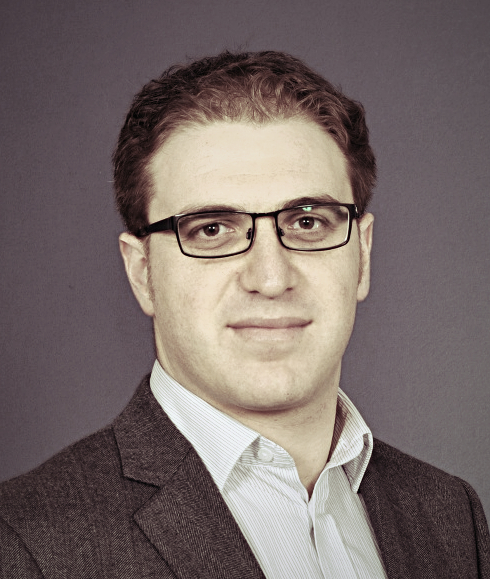}}]
    {Deniz Gündüz }
    Deniz Gündüz [S’03-M’08-SM’13-F’22] received the B.S. degree in electrical and electronics engineering from METU, Turkey in 2002, and the M.S. and Ph.D. degrees in electrical engineering from NYU Tandon School of Engineering (formerly Polytechnic University) in 2004 and 2007, respectively. Currently, he is a Professor of Information Processing in the Electrical and Electronic Engineering Department at Imperial College London, UK, where he also serves as the deputy head of the Intelligent Systems and Networks Group.
\end{IEEEbiography}

\end{document}